\documentstyle[12pt]{article}
\newcommand{\eq}{\begin{equation}}
\newcommand{\en}{\end{equation}}
\newcommand{\BEF}{\begin{figure}}
\newcommand{\EF}{\end{figure}}
\newcommand{\bea}{\begin{eqnarray}}
\newcommand{\eea}{\end{eqnarray}}

\newcommand{\th}{\theta}

\newcommand{\J}{\J_{\mu ab}}
\newcommand{\T}{\T^{\mu\nu}}

\newcommand{\DN}{\not\!\!{D}^{(N)}}
\newcommand{\hDN}{\not\!\!\hat{D}^{(N)}}
\newcommand{\hDk}{\not\!\!\hat{D}^{(k)}}
\newcommand{\hDp}{\not\!\!\hat{D}^{(1)}}
\newcommand{\hD}{\not\!\!\hat{D}}
\newcommand{\hDm}{\not\!\!\hat{D}^{(-1)}}
\newcommand{\hDpm}{\not\!\!\hat{D}^{(\pm 1)}}

\newcommand{\PN}{{P}_{{\rm Ker}~ {\not  D}^{(N)}}}
\newcommand{\hPN}{{P}_{{\rm Ker}~ {\not  \hat D}^{(N)}}}
\newcommand{\Id}{1\!\! 1}

\newcommand{\unity}{1\kern-.65mm \mbox{\form l}}
\newfont{\form}{cmss10}
 
\newcounter{lett2} 
\newenvironment{mathletters}{ 
\refstepcounter{equation} 
\setcounter{lett2}{\value{equation}} 
\setcounter{equation}{0} 
   
}{\setcounter{equation}{\value{lett2}} 
 
\vspace{-.5truecm}\\
 \noindent}

\begin{document}
\font\ninerm = cmr9
\def\footnoterule{\kern-3pt \hrule width \hsize \kern2.5pt}
\setlength{\baselineskip}{13pt}
\begin{flushright}
November 1996\\
MIT-CTP-2578\\
BRX-TH-401
\end{flushright}
\begin{center}
{\Large\bf Non-minimal couplings in two-dimensional gravity: 
a quantum investigation}
\footnote{\ninerm This work is supported
in part by funds provided by the U.S. Department of Energy (D.O.E.)
under cooperative agreement \#DE-FC02-94ER40818,  by NSF grant PHY-9315811, 
and by Istituto Nazionale di Fisica Nucleare (INFN, Frascati, Italy). }\\
\vspace{1cm}
{L. Griguolo$^\dagger$\footnote{e-mail:\tt griguolo@irene.mit.edu}  
    and 
D. Seminara$^\ddagger$\footnote{e-mail:\tt seminara@binah.cc.brandeis.edu}}

\vskip 0.5cm
{\it $^\dagger$ Center for Theoretical Physics\\
Laboratory for  Nuclear Science and Department of Physics\\
Building 6, Massachusetts Institute of Technology\\
Cambridge, Massachusetts 02139, U.S.A.}

{\it $^\ddagger$ Department of  Physics, Brandeis University\\
    Waltham, MA 02254, USA}

\end{center}

\vspace{1cm}
\begin{abstract}
\noindent
We investigate the quantum effects of the non-minimal matter-gravity 
couplings  derived by Cangemi and Jackiw in the realm of a specific 
fermionic theory, namely the abelian {\it Thirring} model on a Riemann 
surface of genus zero and one. The structure and the strength of the 
new interactions are 
seen to be highly constrained, when the topology of the underlying 
manifold is taken into account. As a matter of fact, by requiring to have 
a well-defined action, we are led both to quantization rules for the 
coupling constants and to selection rules for the correlation functions. 
Explicit quantum computations are carried out in genus  one (torus). 
In particular the two-point function and the chiral 
condensate are carefully derived for this  case.  Finally the 
effective gravitational action, coming from integrating out the 
fermionic degrees of freedoom, is presented. It is different from the 
standard Liouville one: a new non-local functional of the conformal factor 
arises and the central charge is improved, depending also on the Thirring 
coupling constant. This last feature opens the  possibility of giving a 
new explicit representation of the minimal series  in terms of a fermionic 
interacting model.
\end{abstract}
\newpage
\section{Introduction}
\baselineskip .8truecm
One of the most intriguing feature of two dimensional quantum field theory is 
the possibility to explore directly  the interplay between geometry, topology 
and purely quantum dynamical effects. In this simplified context one can test, 
for example, the influence of topologically charged fields 
(instantons and monopoles) in gauge theories \cite{Topi} or to study the 
exact dependence of field theoretical quantities from the geometry and the 
topology of the underlying manifold \cite{Geomi}. A more ambitiuos (and interesting) 
task is to study the dynamics of the geometry itself, and particular 
attentions were devoted, in this sense, to non-critical string theory 
\cite{Polli} and stringy-inspired gravities (dilaton gravities) and their 
generalizations \cite{Giacchi}.

\noindent
In ref. \cite{Cangi} Cangemi and Jackiw have shown that  the usual gravity-matter 
interaction in two dimensions can be altered non trivially without  spoiling the 
general covariance or introducing new degrees of freedom.  These novel 
interactions correspond, in a geometric language, to a non-minimal  coupling with
the curvature and to an unconventional one with the volume form.  The first 
addition has been seen to produce, at quantum level,  modifications familiar from
conformal improvements of dynamics, namely a change  in the central charge of the
minimal theory. The second addition is similar  to a costant electromagnetic field 
in flat two-dimensional space-time and reduces to that in absence of curvature.  
In their  investigations the structure of the underlying manifold was not taken 
into account, essentially assuming that the space-time has the topology of the 
plane. 

\noindent
Due to  experience with topological quantization rules \cite{Globi} and 
instanton effects \cite{Raggio} we wonder whether  non-trivial restrictions 
on these couplings arise from the topology  and how their classical and quantum 
dynamics is modified by globality requirements.

\noindent
In this paper, for sake of concretness, we have chosen to investigate 
these issues in the case of the Abelian Thirring model on a Riemann surface
of genus zero and one. Actually the explicit quantal computation of the 
partition function and of some correlators is  carried out in genus one 
(torus), where the topological structure is rich enough to exploit the 
constraints coming from non-trivial homology cycles, flat-gauge connections 
and instanton solutions. (On the sphere the first two features are in fact 
absent). Particular attention is  also devoted to the dynamical consequences 
that these new couplings produce with respect to the usual Thirring model on 
a Riemann surface \cite{Frie}.
 
\noindent
We start, in Sect. 2,  by  briefly reviewing the non-minimal couplings introduced
by Cangemi and Jackiw in ref. \cite{Cangi}. In two dimensions, because of  the 
simplified indeces structure, it is possible to consistently modify the covariant
derivative acting on a field, e.g. on  a Dirac spinor, without introducing new 
degrees of freedom beyond the gravitational ones.  Such modifications induces 
naturally two types of interactions. The first one is an additional 
coupling to the curvature, which, in the case of fermions, simulates the 
conformal improvement  ($R\phi$) usually considered for bosons. The second 
one is, instead, a coupling with the volume two-form of the manifold, which, 
in the flat space-time limit, corresponds  to a sort of background electric 
field. (As we shall see there is indeed a relation between the  $U(1)$-bundle 
of the manifold and this last interaction.)


\noindent
In Sect. 3 we explore the interplay between  a non trivial topology  and
the new couplings, paying particular  attention to the cases of  the 
sphere and  the torus.  The requirement  of having a globally defined 
action induces   quantization rules for the strentgh of these interactions. 
Moreover  we  show that their presence, in general, implies the existence 
of zero modes for the Dirac operator, which, in turn, entails the possibility
of having chirality violating Green functions.

\noindent
However, on the sphere, there is a choice of the coupling constants for 
which the zero modes are absent. In such a case the induced action for 
the conformal factor  displays a new highly non-local contribution apart 
from the  Liouville  one, coming  from the interaction with the volume form. 
On the  other hand  a non vanishing partition function on the torus can be 
only obtained by neglecting the volume-coupling. As a matter of fact  we end 
up with a conventional Liouville action, the original central charge of the 
theory being dressed.


\noindent
In Sect. 4  we investigate the dynamical consequences of the non-minimal
couplings in the contest of a specific interacting fermionic theory,
namely the Abelian Thirring model. We limit ourselves to the torus,
where the algebra is somehow simpler. There we start by observing that 
 the coupling with the volume-form mimics a sort of instanton  background.
By exploiting  this similarity we are able  to write down the generating functional 
in a closed  form  by means of standard technique. However, some subtleties arise 
in the  definiton  of the generating functional, if we want to obtain correlators 
which  are well-defined on the torus.


\noindent
In Sect. 5 we give the explicit form of the two-point function, both in the
theory where the zero modes are absent (i.e. no volume-interaction) and  in 
the theory where they are present. In particular one can easily check 
that a non vanishing fermoinic condensate appears in the last case, as it 
could be expected because of the analogy with the instantons.


\noindent
Let us notice that an interpretation of our result in the spirit of 
quantum field theory at finite temperature is also possible, due to the 
torus topology; we have, for example, discussed the temperature dependence 
of the fermionic condensate as a simple consequence of our computation.

\noindent
Finally in Sect.6 we present our conclusions and the possible extensions 
of this work.

\section{Non-minimal couplings for fermionic theories}
In this section we  briefly  review the non-minimal couplings to gravity
introduced by  Cangemi and Jackiw \cite{Cangi}. However, being interested in the  
abelian {\it Thirring} model,  we  have chosen to  focus our attention only  
on the case of a Dirac spinor  $\psi$,  referring the reader to ref. 
\cite{Cangi} for a more systematic and general presentation.

\noindent
In arbitary dimension  the interaction between gravity and a Dirac spinor 
field $\psi$ is built by introducing a covariant derivative 
\begin{equation}
\label{derivative}
D_\mu\equiv\partial_\mu +\frac{i}{2} \omega^{ab}_\mu ~\sigma_{ab},
\end{equation}
whose main property is to satisfy the algebra
\begin{equation}
\label{algebra}
\left [ D_\mu,D_\nu \right]=\frac{i}{2} R^{ab}_{\mu\nu} ~\sigma_{ab}.
\end{equation}
In eqs. (\ref{derivative}) and (\ref{algebra}) $\sigma_{ab}$ is the 
Lorentz generator corresponding to the given spinor representation,
while $R^{ab}_{\mu\nu}(\equiv\partial_\mu 
\omega^{ab}_\nu+\omega^a_{~c\mu} \omega^{cb}_\nu-\mu\leftrightarrow\nu)$ 
is the curvature two-form.

\noindent
In ref. \cite{Cangi} the basic idea is to modify the algebra (\ref{algebra})
as follows
\begin{equation}
\label{algebra1}
\left [ D_\mu,D_\nu \right]=\frac{i}{2} R^{ab}_{\mu\nu} ~\sigma_{ab}
+i  {\cal F}_{\mu\nu} \Id,
\end{equation}
where ${\cal F}_{\mu\nu}$ is a two form. In dimensions greater than two, 
this new contribution  cannot be  generated  only by means  of the 
gravitational variables; it arises  when  electromagnetic 
deegres of  freedom are dynamically  active -- but here we do not include   
these  additional variables. In two dimensions, however,  
gravitational variables allow constructing the required  form; in fact 
we can take
\begin{equation}
\label{pipino}
{\cal F}_{\mu\nu}={\cal F}\sqrt{g}~\epsilon_{\mu\nu},
\end{equation}
where $g$ is the determinant of the (Riemanian) metric of the manifold in 
consideration, $\epsilon_{\mu\nu}$  the ordinary Levi-Civita tensor and 
finally ${\cal F}$  any  scalar field built out from the metric [ e.g. any 
function of the scalar curvature $R(\equiv e^\mu_a e^\nu_b R^{ab}_{\mu\nu}
$)]. 

\noindent
Following ref. \cite{Cangi}, in the sequel of this paragraph we shall only 
consider  the two simplest contributions to ${\cal F}$, i.e.
\begin{equation}
\label{po5}
{\cal F}=A~ R+B,
\end{equation}
with $A$ and $B$ two  constants  setting  the strength of the new 
interactions. Moreover, in ref. \cite{Cangi} this particular choice 
was also seen to fit very naturally  into a gauge theoretical 
description  of the two-dimensional gravity in terms of the extended 
Poincar\`e group \cite{Cangi1}.

\noindent
As it is clear from the aforementioned analogy with the gauge field, 
the new terms in eq. (\ref{algebra1}) can be originated by adding 
suitable  vector potentials  to the covariant derivative (\ref{derivative}). 
The  contribution  proportional to the curvature is obviously generated by the 
combination  $i A \omega_\mu$ ($\omega_\mu\equiv 1/2 
\epsilon_{ab} \omega^{ab}_\mu$), while the other one involves some subtleties.

\noindent 
The two-form  ${1\over 2} B \sqrt{g} \epsilon_{\mu\nu} dx^\mu\wedge 
dx^\nu$ is proportional to 
the volume form ${\cal V}
(\equiv 1/2\sqrt{g} \epsilon_{\mu\nu} dx^\mu\wedge  dx^\nu)$. Since 
${\cal V}$  is closed we can define  locally a one-form $a$ for which
\begin{equation}
\label{Vollo}
\partial_\mu a_\nu-\partial_\nu a_\mu-B \sqrt{g}~ \epsilon_{\mu\nu}=0,
\end{equation}
where  $a$ is  determined up to an exact (globally defined) one-form 
$d\beta$
\begin{equation}
\label{gauge}
a_\mu\rightarrow a_\mu+\partial_\mu\beta.
\end{equation}
Thus this interaction can be obtained by adding the combination 
$i  a_\mu$ to the covariant derivative (\ref{derivative}).
The improved action for a Dirac spinor field  now  reads
\begin{equation}
\label{oppio}
S=\int d^2x~\det(e)~ \bar\psi~ e^\mu_a~\gamma^a i \left
(\partial_\mu+ \frac{i}{2} \omega_\mu~\gamma_{5}+ i A\omega_\mu 
+ i  a_\mu\right )\psi- \int d^2 x \epsilon^{\mu\nu}\rho\left (
\partial_\mu a_\nu-\frac{B}{2}\epsilon_{ab} e^a_\mu e^b_\nu\right )
\end{equation}
where $\rho$ is a Lagrange multiplier which enforces the constraint
(\ref{Vollo}) and we have used the explicit form of $\sigma_{ab}=
\frac{i}{4}\epsilon_{ab} \gamma_5$ in  two dimensions.
	
\noindent
The invariance of such an action under diffeomorphism is manifest while 
local $SO(2)$ is  preserved  only if we modify the transformation law 
for the Dirac  fields as follows
\begin{equation}
\label{pipino2}
\psi\rightarrow \exp({i\over 2}\alpha\gamma_{5}+i A\alpha)\psi,
\,\,\,\,\,\,\,\,,\,\,\,\,
\bar{\psi}\rightarrow \bar{\psi}\exp(-{i\over 2}\alpha\gamma_{5}-i A\alpha).
\end{equation}
Moreover the presence of the field $a_\mu$ introduces a further $U(1)$
symmetry. In fact if we change the fields according to
\begin{equation}
\label{pipino1}
\psi\rightarrow \exp(i\beta)\psi, \,\,\,\,\,\,\,\,\,\,\,\,
\bar{\psi}\rightarrow \bar{\psi}\exp(-i\beta)
\,\,\,\,\,\,\,\,\,\,\,\ {\rm and} \ \ a_\mu \rightarrow a_\mu+
 \exp(-i \beta)\partial_\mu \exp(i \beta),
\end{equation}
the action is invariant. At this level we can consider $a_{\mu}$ as an 
ordinary $U(1)$ connection.

\noindent
Some comments about the new interactions appearing in the Lagrangian 
(\ref{oppio}) are now  in order. The origin of the coupling 
$A \omega_\mu$ can be traced back to the abelian character of the 
rotation group in two dimensions. In fact in this case we are not 
forced to represent our group with traceless generator. 
[This is analogous to what happens in abelian gauge theory, where one 
can arbitrarily mix the vectorial and axial coupling without problems. Let us recall,
on the other hand, that this is impossible in the non-abelian case.] 
Moreover the introduction of the $A\omega_\mu$ term 
changes substantially the conformal structure of the theory. The 
conformal invariance of the {\it Dirac} part of the action is preserved
only if we modify the transformation for the spinors as follows (we 
call $\sigma$ the parameter of the conformal transformation)
\begin{equation}
\psi(x)\rightarrow \exp\left[\left(-\frac{\Id}{2}+A \gamma_5\right)
\sigma(x) \right ]\psi(x),
\end{equation}
and similarly for $\bar \psi(x)$. In  other words the right and left 
component of the spinor have now different conformal weights.

\noindent
The group theoretical  root of the interaction with the voulme 
potential $a_\mu$ is, instead, located in the possibility of 
modifying the algebra of traslations with a central extension, 
namely
\begin{equation}
[P_{a},P_{b}]=\epsilon_{ab}I.
\end{equation}
Such extension arises naturally in two dimensions. A classical
example is the case of the Schr\"odinger equation coupled to a
costant magnetic field.

\noindent
On a compact  Riemannian manifold both $\omega_\mu$ and $a_\mu$
possess, in general, a non trivial winding number, i.e.  they 
do not exist as 1-forms on the manifold, but only as connections
on the tangent and $U(1)$ bundle respectively. In the following  
we shall see how this fact implies a certain number of non trivial
constraints  on the couplings, the partition function and the 
correlators of our fermionic model.


\section{Non-minimal couplings on the sphere and on the torus}
The abelian {\it Thirring} model in the presence of the non-minimal couplings
is immediately obtained by adding the   usual current-current interaction 
to the improved Lagrangian (\ref{oppio}). However, before committing  ourselves
with this specific fermionic theory, we shall address the preliminary question 
of defining the Dirac operator in eq. (\ref{oppio}) on manifolds with non trivial
topology, specifically the sphere and the torus. This analysis  will lead us to 
single out some kinematical constraints on the coupling costants and the Green
functions of our theory. 

\noindent 
At the end of this section we will also spend some words  about 
what  results might  be taken over directly to higher genus.  


\noindent

\subsection{The sphere}
The geometry of $S^2$ is described by the zweibein: 
\begin{equation}
e^{a}_{\mu}=e^{\sigma}\hat{e}^{a}_{\mu},
\label{s1}
\end{equation}
$\hat{e}^{a}_{\mu}$ being a reference zweibein for a two-sphere of costant 
radius (taken equal to $1$), while $\sigma$ is the conformal factor, which 
will represent the Liouville mode in the following. 

\noindent
Choosing angular 
coordinates $(\theta,\varphi)$, $0\leq\theta<\pi\, , \,0\leq\varphi<2\pi$ 
and fixing the Lorentz frame we get for $\hat{e}^{a}_{\mu}$:
\begin{equation}
\hat{e}^{a}_{\mu}=
\left (
\begin{array}{cc}
\cos\varphi & -\sin\varphi\sin\th\\
\sin\varphi & \cos\varphi\sin\th \end{array}\right).
\label{s3}
\end{equation}
The metric related to the zweibein in eq.  (\ref{s1}) is therefore 
$g_{\mu\nu}=\delta_{ab}e^{a}_{\mu}e^{b}_{\nu}=e^{2\sigma}\delta_{ab}
\hat e^{a}_{\mu}\hat e^{b}_{\nu}=e^{2\sigma}\hat{g}_{\mu\nu}$. 

\noindent
In order to construct the non-minimal couplings we have, first of all, to 
obtain a solution for the equation (\ref{Vollo}), defining the potential 
for the volume-form. To this purpose we notice that on $S^2$ we have 
instanton fields  solving the Maxwell equations
\begin{equation}
\nabla_\mu F^{\mu\nu}=0.
\end{equation}
The most general form of such solutions, with regular behaviour, 
is given by
\begin{equation} 
\label{frau}
F^{(k)}_{\mu\nu}\equiv\partial_\mu A^{(k)}_\nu-\partial_\nu A^{(k)}_\mu=
\frac{2\pi k}{V}\sqrt{g}~ \epsilon_{\mu\nu} \ \ \ \
\  {\rm with} \ \ k\in Z,
\end{equation} 
where $V$ is the volume of the manifold and $k$ is the winding number 
\begin{equation}
k=\frac{1}{4\pi}\int d^2x~ \epsilon^{\mu\nu}~F^{(k)}_{\mu\nu}. 
\end{equation}
The corresponding gauge-connection can be now written (in a single patch)
\begin{equation}
\label{instanton2}
A^{(k)}_\mu= \hat A^{(k)}_\mu -\pi k~ \eta_{\mu}^{~\nu}
\partial_\nu \lambda,
\end{equation}
where $\eta_{\mu\nu}\equiv \sqrt{g}~\epsilon_{\mu\nu}$ and 
$\lambda$ solves the equation
\begin{equation}
\sqrt{g}~\triangle \lambda=\frac{\sqrt{g}}{V} -\frac{\sqrt{\hat
g}}{\hat V}.
\label{nonloc}
\end{equation}
The field $\hat A^{(k)}_\mu $ in eq. (\ref{instanton2}) is the instanton 
computed in the background geometry $\hat{g}_{\mu\nu}$. In this way 
$\lambda$ carries the dependence on the conformal factor, while 
$\hat A^{(k)}_\mu $ modulo gauge transformation is 
(in angular coordinates)\footnote{We recall that only the one 
form $(\sin\theta~ d\phi)$ has a global meaning on the sphere, 
while $d\phi$ does not.}
\begin{equation}
\hat A^{(k)}=\hat A^{(k)}_{\varphi}~ (\sin\theta~ d\varphi)=
{k\over 2}\tan{\th\over 2}~ (\sin\theta d\varphi).
\label{s12}
\end{equation}
The expression (\ref{s12}) for the instanton solution has, obviously, 
only a local meaning due to the  singularity at $\th=\pi$, present in 
$\hat A^{(k)}_\varphi $. Nevertheless  $\hat A^{(k)}_\mu $ is a connection 
on a non-trivial $U(1)$-bundle; in fact we need at least two patches to cover 
the sphere and in the overlapping region non-trivial connection are related by 
a gauge transformation. Describing  the system in a patch where $\hat A^{(1)}_\mu$ 
is  regular at $\th=\pi$ we get \cite{Gri}
\begin{equation}
\hat {A'}^{(k)}=\hat {A'}^{(k)}_{\varphi} (\sin\theta d\varphi)
=-{k\over 2}\cot{\theta\over 2}  (\sin\theta d\varphi)
\label{s13}
\end{equation}
the two expression being connected  by the 
gauge transformation
\begin{equation}
\hat A^{(k)}-\hat{A'}^{(k)}=k{d\varphi}={i}\,g^{-1} d g,
\label{s18}
\end{equation}
where $g=\exp[-i k\varphi]$ is a well-defined $U(1)$-valued function on 
the overlapping region. 

\noindent
Now, by exploiting the resemblance between eq. (\ref{frau}) and (\ref{Vollo}), 
we can immediately write down  an expression for the volume-potential
\begin{equation}
a_{\mu}={BV\over 2\pi}A^{(1)}_{\mu}.
\label{pipino4}
\end{equation}
\noindent
The non trivial structure of the instanton solution is obviously
inherited by the volume potential $a_\mu$. In fact, when changing patch, 
one gets as a consequence of (\ref{s18}) 
\begin{equation}
{a}-{a'}=\frac{BV}{2\pi}{d\varphi}=i
\exp[i{BV\over 2\pi}\varphi]~ d ~\exp[-i{BV\over 2\pi}\varphi].
\label{sbis18}
\end{equation}
[Eq. (\ref{sbis18}) seems to suggest that $a_\mu$ is a gauge
connection in a non trivial $U(1)$ bundle and this would imply 
that
\begin{equation}
{BV\over 2\pi}=n,\ \ \ \ \ \ {\rm with}\ \  n \in Z.
\label{s20}
\end{equation}
Actually this conclusion is somehow misleading. In fact, as we
shall see  in the following, the requirement (\ref{s20}) is too
strong.]

\noindent 
On the sphere also the spin-connection shares an analogous  non
trivial structure. Due to the fact that the Euler mumber 
$\chi$ is not zero, $\chi=2$, we expect that the  spin-connection 
$\omega_{\mu}$ does not have a global definition. Actually in the parametrization  
we have choosen the spin-connection can be written as \cite{Gri}
\begin{equation}
\label{spinconn}
\omega_\mu= \hat A^{(2)}_\mu -\eta_{\mu}^{~\nu} \partial_\nu \sigma.
\end{equation}
The relevant gauge-transformation, connecting $\omega_\mu$ in the two 
overlapping  patches, turns out to be in this case  $g=\exp[-2 i\varphi]$. 
[ A consideration similar to the one presented above holds also here.]

\noindent
In order that the quantum theory be  defined, what we, really, need
is that the {\it Dirac} operator appearing in eq. (\ref{oppio}) has 
a global meaning, i.e. its eigenvalue problem must be patch-independent.
This is simply obtained if the {\it Dirac} operator
\begin{equation} 
D=i\gamma^\mu\left[\partial_\mu +i\left(\frac{1}{2} \gamma_5+A\Id\right ) 
\omega_\mu + a_{\mu}\right ]
\end{equation}
undergoes through a similitude transformation $D'=U^{-1}DU$ 
under change of patch. Using the previous result, one can promptly 
check that such an $U$ exists and it is given by
\begin{equation}
U=\exp\left[-i\left ({\gamma_5}+ \left(2 A+\frac{BV}{2\pi}\right)\Id\right)
\varphi\right ].
\end{equation} 
The requirement that $U$ is well-defined in the overlapping region leads to 
the following quantization condition
\begin{equation}
\left(2 A+\frac{BV}{2\pi}\right)=N \  \ \ \ {\rm with}\ N\in \ Z.
\label{vcv}
\end{equation}
Eq. (\ref{vcv}) has a simple geometrical intepretation in the framework of 
the bundle geometry. Neither $a_\mu$ neither $\omega_\mu$ exist separately 
as connection on the principal bundle, but only the combination $a_\mu
+A \omega_\mu$ has a global meaning. 

\noindent
This conclusion becomes even more transparent, if we look at the problem
from the gauge theoretical point of view developed by Cangemi and Jackiw 
\cite{Cangi}. In this approach $\omega_\mu$ and $a_\mu$ are not separate 
objects but they form, with  $e^a_\mu$, a gauge connection living in the 
principal-bundle of the central extended Poincar\`e group. Thus it is only 
this quantity that must exist globally and not its components.  One can
check that this requirement is equivalent to the condition found before.
[Actually a complete group theoretical analysis in terms of the extended 
Poincar\`e group is quite tricky  because it involves infinite dimensional 
representation of the relevant Lie algebra \cite{Cangi}, while in terms 
of the  properties of the {\it Dirac} operator is quite 
straightforward.]
 
\noindent
The next step in our analysis is to explore the influence  of these new 
interaction on the Dirac operator
\eq
\hat{D}^{(N)}=\,\gamma_{a}e^{\mu}_{a}[\,iD_{\mu}+A\omega_{\mu}+
a_{\mu}\,],
\label{s50}
\en
whose spectrum governs the quantum dynamics of our model. 

\noindent
To every eigenfunction  $\Psi_{i}$ of the operator $\hat{D}^{(N)}$ 
with a non-zero eigenvalue $E_{i}$ another eigenfunction $\Psi_{-i}=
\gamma_{5}\Psi_{i}$ corresponds, with eigenvalue $E_i=-E_{-i}$. 
For non-vanishing $N$,  zero modes are present and they  have definite 
chirality  i.e. \[ \gamma_{5}\chi^{(N)}_{i}=\pm \chi^{(N)}_{i}.\] 
Their  number is $N=n_{+}-n_{-}$, $n_{+}$ corresponding  to positive chirality
and $n_{-}$ to the negative 	one \cite{Atty}:
\begin{equation}
\begin{array}{llllll}
n_{+}=0   & &  n_{-}=|N| & &  &  N\geq 0\\
n_{+}=|N| & &  n_{-}=0   & &  &  N\leq 0
\end{array}.
\label{s51}
\end{equation}
\noindent
Due to the presence of zero-modes, for generic $N$, the partition function 
vanishes: the generating functional is nevertheless not  zero, and 
chirality-violating correlation functions, with chirality selection 
rules governed by $N$, appear on the theory.  In fact, using the 
notation
\begin{equation}
\psi=\left(
\begin{array}{c} 
\psi_R\\ 
\psi_L
\end{array}
\right ),
\ \ \ \ \ \ \
\bar \psi=\left(\bar\psi_R, \bar\psi_L\right ),
\end{equation}
one can check that the only non-vanishing correlation functions are
\begin{equation}
\label{giga}
<\bar\psi_R(x_1) \psi_R (y_1)......\bar\psi_R(x_p) \psi_R (y_p)
\bar\psi_L(w_1) \psi_L (z_1)......\bar\psi_L(w_q) \psi_L (z_q)>,
\end{equation}
with $p-q=N$. An explicit  proof of this statement will be given in sect.
4, where we construct the generating functional for the Thirring model 
on the torus. However it is easy to realize that this property is independent
of the detail of the model under scrutiny. 

\noindent
Alternatively, in order to obtain a non-trivial partition function, we can 
choose $A$ and $B$ to give $N=0$. 
A non-trivial modification of the usual Liouville dynamics, for genus-zero 
Riemannn surface, is therefore produced by the presence of the field 
$\lambda$ linked to the volume-potential. (We stress that it is not an 
independent field but a non-local functional of the Liouville mode). 
The partition function for a Dirac fermion is
\begin{equation}
{\cal Z}=\int Dg_{\mu\nu}D\bar{\psi}D\psi\exp[-\int d^2x\sqrt{g}
\bar{\psi} D^{(0)}\psi],
\label{party}
\end{equation}
and performing the functional integration over the fermions field we get
\begin{equation}
\label{rty}
{\cal Z}=\int Dg_{\mu\nu}\exp[-S_{Liou.}(\sigma,\hat{g}_{\mu\nu})]
\exp[-S_{n}(\sigma,\lambda,\hat{g}_{\mu\nu})],
\label{party1}
\end{equation}
where $S_{Liou.}$ is the usual Liouville action and  the new contribution 
coming from the non-minimal coupling being
\begin{equation}
\label{addition}
S_{n}(\sigma,\lambda,\hat{g}_{\mu\nu})={N^2\pi\over 2}
\int d^2x\sqrt{\hat{g}}(\lambda-{\sigma\over 2\pi})\hat{\triangle}
(\lambda-{\sigma\over 2\pi}).
\label{party2}
\end{equation}
The $\lambda-$dependence of this additonal  contribution can be traced back 
to the fact that the constraint defining the vector potential $a_\mu$  is 
not conformal invariant. Moreover we notice that the combination $-N^2/8\pi
\sigma\hat \triangle\sigma $, due to the coupling $A\omega_\mu$, improves 
the central charge of the theory.

\noindent
Exploring  how this new term affects the standard Liouville dynamics is 
beyond the aim of the present paper, even if it seems reasonable to expect 
substantial changes in the quantum dynamics of gravity.

\subsection{The torus}
With a suitable choice of the coordinates and tangent frame we can always
write the field $e^a_{\mu}$, describing a generic torus geometry, 
in the form $e^a_{\mu}={\rm e}^{\sigma }\hat e^a_{\mu}$ where
\begin{equation}
e^a_\mu=e^{\sigma} {\hat e}^a_\mu=e^{\sigma}\left (
\begin{array}{cc}
\tau_2 & \tau_1\\
0 & 1 
\end{array}
\right),
\label{eq1}
\end{equation}
$\tau=\tau_1+ i \tau_2$ being  the Teichmuller parameter. In eq. (\ref{eq1}) 
the fundamental region for the coordinates has  taken to be the square  
$0\le x^1, x^2  <1$, while $\sigma$ is  the conformal factor. The ensuing
metric is obviously $g_{\mu\nu}=\delta_{ab}e^{a}_{\mu}e^{b}_{\nu}=e^{2\sigma}
\delta_{ab}\hat e^{a}_{\mu}\hat e^{b}_{\nu}=e^{2\sigma}\hat{g}_{\mu\nu}$. Let 
us remark that, at variance with the case of the sphere, the background metric 
is flat. 

\noindent
The presence of the Teichmuller parameter is not the only new feature of
the torus. In fact, due to the existence of non trivial cycles, it admits 
more than one spin structure. Such structures correspond to the 
four possible boundary conditions, which  one can choose for the 
fermions, when  $x^1\to x^1+1$ or $x^2\to x^2+1$, namely $(A,A)$, 
$(P,A)$, $(A,P)$ and  $(P,P)$, where $P(A)$ stands for periodic 
(antiperiodic). 
Here we will  be only concerned with the (A,A) boundary conditions
even if the invariance under global diffeomorphism  would requires 
to sum over all of them. By the way, this choice  will allow
a  finite-temperature  interpretation of our results.

\noindent
We have seen in the previous subsection how the volume potential can be 
obtained 
from a connection that belongs to a non-trivial $U(1)$-bundle over the 
Riemann surface and solves the Maxwell equations. 
Let us start therefore with discussing the structure of the non-trivial 
$U(1)$-bundle on a two-dimensional compact euclidean torus. The same 
procedure we have followed for the two-sphere leads to write the relevant 
gauge-connection (in a single patch)
\begin{equation}
\label{instantont}
A^{(k)}_\mu= \hat A^{(k)}_\mu -\pi k \eta_{\mu}^{~\nu}
\partial_\nu \lambda+{2\pi}\alpha^{(k)}_\mu,
\end{equation}
where $\alpha^{(k)}_\mu$ is a flat connection (i.e.an harmonic one form 
$\epsilon^{\mu\nu}\nabla_\mu \alpha^{(k)}_\nu=0)$, while $\lambda$ 
is defined as in eq. (\ref{nonloc}). 

\noindent
We notice that on the torus, at variance with the genus-zero case, 
we have family of gauge inequivalent solutions of the Maxwell equations, 
labelled by the choice of the flat connection $\alpha^{(k)}_{\mu}$ in 
eq. (\ref{instantont}). It cannot be removed by a well-defined gauge 
transformation: we are only allowed to perform 
\begin{equation}
\alpha_{\mu}\rightarrow \alpha_{\mu}+n_{\mu},\,\,\,\,\,
n_{\mu}=(n_{1},n_{2}) 
\label{large}
\end{equation}
$n_{1},n_{2}\in Z$, using as gauge function 
$g=\exp[i{2\pi} n_\mu x^{\mu}]$, 
that is single-valued on the torus. 
These transformations are not connected to the 
identity and are usually called {\it large} gauge transformations. In 
the standard $U(1)$-gauge theory, where gauge invariance is respected, 
eq. (\ref{large}) implies that the flat-connections $\alpha_{\mu}$ take 
values in the real interval $[0,1]$. The existence of gauge inequivalent 
solutions to the Maxwell equations reflects, in our case, in the 
appearence of a new (global) degree of freedom inside the volume-potential. 

\noindent
The field $\hat A^{(k)}_\mu $ in eq. 
(\ref{instantont}) is the instanton computed in the background 
given by $\hat g_{\mu\nu}$. In this way  $\lambda$ carries the 
dependence on the conformal factor, while  $\hat A^{(k)}_\mu$ 
depends only on the moduli space, described by $\tau$.
On the torus   $\hat A^{(k)}_\mu$, modulo gauge transformations, is
\begin{equation}
\hat A^{(k)}_\mu=-\frac{\pi k}{\hat V} \sqrt{\hat g}
\epsilon_{\mu\nu} x^\nu=-{\pi k}{} 
\epsilon_{\mu\nu} x^\nu.
\end{equation}
Being  a connection with a non-vanishing winding number, 
$A^{(k)}_\mu$ has  non-trivial boundary conditions after 
a homology cycle. Explicitly they read
\begin{mathletters}
\begin{eqnarray}
&&A_\mu^{(k)}(x^1+1,x^2)=A_\mu^{(k)}(x^1,x^2)+\partial_\mu\left ({\pi
k} x^2\right) \\
&&A_\mu^{(k)}(x^1,x^2+1)=A_\mu^{(k)}(x^1,x^2)-\partial_\mu\left (\pi
k x^1\right). 
\label{contorno}
\end{eqnarray}
\end{mathletters}  

\noindent
As in the $S^2$ case we have:
\begin{equation} 
\label{gghh}
a_{\mu}={BV\over 2\pi}A^{(1)}_{\mu}.
\end{equation}
As before $a_\mu$ inherits non-trivial boundary conditions after a homology 
cycle from the instanton field. On the other hand the Euler number of the 
torus is zero, therefore the tangent-bundle is trivial  and no global
issue seems to be raised by $\omega_{\mu}$.

\noindent
We observe that, due to the presence of large gauge transformations, the 
parameter $B$ has to respect two constraints in order to satisfy the global 
definition of the theory: 
$g\!=\!\exp[i B V g^{\mu\nu}n_{\mu}x_{\nu}]$ has to be well defined on 
the torus, while the now $B$-dependent transition functions $g_{ij}$ 
for the Dirac operator, between different patches $U_i,U_j$, 
must obey to the consistency condition $g_{ij}=g_{ik}g_{kj}$ \cite{Libro}. One 
can explicitly check that both requirements are satisfied by choosing 
\begin{equation}
\label{bibbolo1}
{BV\over 2\pi}=N,
\end{equation}

\noindent
As a consequence of the relation between the instanton bundle and the 
volume form $a_\mu$, the spinor-bundle also  possesses non-trivial  boundary  
conditions after the an homology cycle, we have in fact 
\begin{mathletters}
\label{bondo1}
\begin{eqnarray}
&&\psi(x^1+1 , x^2)=\exp \left ( i\pi (1+ N x^2)\right)\psi (x^1, x^2), \\
&&\psi(x^1, x^2 +1)=\exp \left (i \pi(1- N x^1) \right)\psi (x^1, x^2),
\end{eqnarray}
\end{mathletters}  
where the antiperiodic boundary conditions has been taken into account. [ Let us 
warn the reader that this feature will have some subtle implications at the quantum 
level as we shall see in the next sections.]


\noindent

\noindent
We end the subsection by remarking that in the case of the torus, for 
$B\neq 0$, the Dirac operator has always zero-modes: due to the triviality 
of the tangent-bundle we cannot match the effect of the $A$-coupling with 
the $B$-coupling to have a partition function different from zero.  Thus the  
theory on the torus will always  produce, for $B\neq 0$,  Green function in  
the  chirality-violating sector, in strict analogy with  Bardacki-Crescimanno 
model \cite{Bardi}.  For $B=0$, i.e. only the $A\omega_\mu$ coupling is 
present, we  have a partition  function different from  zero. However it is 
the standard Liouville action with an improved central charge given by $c=1-
12 A^2$.

\noindent
Some comments about the possible extension of the previous analysis to higher
genus   are now in order. It is clear that the procedure to construct the
explicit expression for the volume form is independent of the genus. In
fact it relies only upon some general properties of the $U(1)-$bundle.
Moreover in this language the quantization rules for the $(A,B)-$coupling 
expresses the fact that the  Chern-class associated to the connection 
$A\omega_\mu+ B a_\mu$ must be an integer:
\begin{equation}
\frac{A}{2\pi}\int d^2x~ \epsilon^{\mu\nu}~\partial_{\mu}\omega_{\nu}+
\frac{B}{2\pi}\int d^2x~ \epsilon^{\mu\nu}~\partial_{\mu}a_{\nu}=
(2-2 g) A+ \frac{B V}{2\pi}=N,
\end{equation}
where $g$ is the genus of the manifold. For $g=0$ and $1$, this condition
reduces to the ones found above. 

\section{The abelian Thirring model with the non minimal couplings}

We are now ready to discuss the dynamical consequences of the non-minimal 
gravitational couplings in a specific interacting fermionic model, namely
the Abelian {\it Thirring} model \cite{Thir}. In curved space the theory 
has received a certain amount of attentions in the past years: 
stringy-like applications (the so-called Thirring strings \cite{Baggi}) 
were discussed in connection with a spontaneous symmetry-breaking mechanism. 
The partition function for the minimal-coupled model was computed 
on a generic Riemann surface in \cite{Frie} and on a cylinder with twisted 
boundary condition \cite{Destri}. A formal computation of the propagator was 
also 
performed in \cite{Bote}. In the following we restrict ourselves to the case 
of the torus, generalizing the results of \cite{Frie} and \cite{Bote} to 
the non-minimal coupled case in genus one where new features arise. 
Actually it is not difficult to extend our calculations in higher genus, by 
using the well-known properties of theta-functions on Riemann surfaces, but we 
have choosen to limit ourselves to the torus case for a double reason: the 
computations are more explicit without loss of physical consequences and 
the finite temperature interpretation of the theory, 
after a partial decompactification, is also possible. In this sense our results 
can be related with the ones presented in \cite{Wippi}, where gauged and 
un-gauged Thirring models on the torus were studied.
fiore
\noindent
On a two-dimensional euclidean torus the Abelian {\it Thirring} model with 
non-minimal gravitational couplings is characterized by the classical 
Lagrangian
\begin{equation}
\label{Lagrangiantre}
{\cal L}=
\bar\psi\gamma^\mu(i\nabla_\mu + A \omega_{\mu}+
a_{\mu})\psi+\frac{g^2}{4}
\bar\psi\gamma^\mu\psi\bar\psi\gamma_\mu\psi
-\epsilon^{\mu\nu}\rho(
\partial_\mu a_\nu-\frac{B}{2}\epsilon_{ab} e^a_\mu e^b_\nu),
\end{equation} 
where the couplings of the new interactions are subject to the constraints
(\ref{bibbolo1}).

\noindent
An action quadratic in the spinor field, which is more  suitable
for the path integral formalism, can be achieved by introducing
a vector field $H_\mu$. The lagrangian now reads
\begin{equation}
\label{Lagrangianquattro}
{\cal L}=
\bar\psi\gamma^\mu \left (i\nabla_\mu 
+A \omega_{\mu}+a_{\mu}-g H_\mu \right )\psi
+g^{\mu\nu} H_\mu H_\nu -\epsilon^{\mu\nu}\rho(
\partial_\mu a_\nu-\frac{B}{2}\epsilon_{ab} e^a_\mu e^b_\nu).
\end{equation}    
In the Lagrangian (\ref{Lagrangianquattro}) the fermionic sector is now also
invariant under the gauge transformation
\begin{equation}
\label{transform}
\psi(x)\to u(x)\psi(x)\ \ \ \ \ \ \bar \psi(x)\to \bar \psi(x) u^{-1}(x)
\ \ \ \ \ \ \ {\rm  and}\ \  H_\mu(x)\to H_\mu(x)+u^{-1}(x)\partial_\mu u(x),
\end{equation}
where $u(x)$ is an element of $U(1)$. This symmetry is, however, 
explicitly broken by the quadratic piece in $H_{\mu}(x)$. As a matter of fact 
$H_{\mu}$ is not a gauge connection, and consequentely its harmonic piece 
$h_{\mu}$ is not restricted to take values in $[0,1]$. 

\noindent
The quantum version of the abelian Thirring model in the presence of 
the new  gravitational couplings is defined by the following generating 
functional
\begin{eqnarray}
\label{Generat}
&&\!\!\!\!\!\!\!\!
Z[\eta,\bar\eta]=\!\!\!
\int{\cal D}\rho  {\cal D}a_\mu {\cal D}\bar\psi{\cal D}\psi {\cal D} H_\mu
\exp \Biggl[\int d^2x \det(e) 
\Biggl (\bar\psi \gamma^\mu [i\nabla_\mu +A \omega_\mu
+a_{\mu}-g H_\mu] \psi+g^{\mu\nu}H_\mu H_{\nu}\nonumber\\
&&\!\!\!\!\!\!\!\!
-\epsilon^{\mu\nu}\rho(
\partial_\mu a_\nu-\frac{B}{2}\epsilon_{ab} e^a_\mu e^b_\nu)+
+\bar\eta\exp\left (i \int_{\gamma_{Ox}}  a_{\mu} dx^\mu\right )
\psi+
\bar\psi\exp\left (-i \int_{\gamma_{Ox}} a_\mu dx^\mu\right )
\eta\Biggr )\Biggr ],
\end{eqnarray}
being $\eta,\bar{\eta}$ the fermionic sources and  $\gamma_{Ox}$ the  
geodesic connecting  $x$ with a given reference point $O$.  

\noindent 
Note that the requirement that correlation
functions be  single-valued when  changing  patches has forced us to 
introduce an extra 
phase factor in the coupling between the external current and the  
fermionic field in  eq. (\ref{Generat})\footnote{Actually one should 
introduce an  analogous phase depending on the spin connection 
$\omega_\mu$.  This additonal phase would take care of the possible 
Lorentz  tansformation when one changes the patch. Since  geometry is  
a fixed background,  this extra contribution is only an overall factor, 
which can be diregarded for the moment and  eventually restored at the 
end of the computation.}. The presence of this  additional contribution 
takes care of the non trivial boundary conditions that the spinor-bundle 
possess as a consequence of  the fact that $a_\mu$ is a connection with 
a non vanishing  winding number. [ See the subsec. about the torus.] 
Moreover the correlation functions will be automatically invariant under 
the $U(1)$ transformations.

\noindent
A puzzling feature  might be, obviuosly,  the dependence of our results on 
the reference point $O$ and the  curve $\gamma_{Ox}$, used to define the 
generating functional (\ref{Generat}). As we shall  see in the following,  
though this dependence exists, it is quite trivial  and does not affect  
the physical content of the theory.

\noindent
Performing the integral over $\rho$ the field $a_\mu$  in eq. 
(\ref{Generat}) is replaced  by the expression (\ref{gghh}). Moreover 
the integration over $a_\mu$ reduces to an ordinary integral over all 
the gauge inequivalent flat connections $\alpha_\mu$, which parameterizes 
the moduli space of the solutions of the constraint (\ref{Vollo}) 
(for details see the previous subsection). 
Therefore the  generating functional now reads
\begin{eqnarray}
\label{Generat1}
&&Z[\eta,\bar\eta]=\!\!\int^1_0 \!\!\!d \alpha_\mu\!\!\int\!\!
{\cal D}\bar\psi{\cal D}\psi {\cal D} H_\mu \exp\Biggl[\int d^2x \det(e) 
\Biggl (\bar\psi \gamma^\mu [i\nabla_\mu +A \omega_\mu
+N A^{(1)}_\mu -g H_\mu] \psi\nonumber\\
&&+g^{\mu\nu}H_\mu H_{\nu}+\bar\eta\exp\left (i N\int_{\gamma_{Ox}} 
A^{(1)}_{\mu} dx^\mu\right )\psi+
\bar\psi\exp\left (-i N \int_{\gamma_{Ox}} A^{(1)}_\mu(x) dx^\mu\right )
\eta\Biggr )\Biggr ],
\end{eqnarray}
where we have used that $B V/2\pi$ is an integer $N$ and the Landau gauge 
for the field $a_{\mu}$

\noindent

\noindent
We are ready now to evaluate  the fermionic integral in 
eq. (\ref{Generat1}). 
In the following we shall use a compact notation to avoid cumbersome 
expressions. The symbol $\not\!\!{D}^{(N)}$ will stands for
\begin{equation}
\gamma^\mu \left(i\nabla^x_\mu + A\omega_{\mu}+N A^{(1)}_\mu
-g H_\mu\right ),
\end{equation}						
while $\eta_a\ (\bar \eta_a)$ will denote the external currents
dressed by the corresponding phase factor. In addition we shall 
introduce a new operator $S^{(N)}$ defined by
\begin{equation}
S^{(N)}\circ \DN=\DN\circ S^{(N)}= {1\!\! 1} 
-{P}_{{\rm Ker}~ {\not  D}^{(N)}}.
\end{equation}
Here $\PN$ is the self-adjoint projector on the Kernel of the 
operator $\DN$. Recall, infact, that the Dirac operator has 
zero modes for $ N\not = 0$; as matter of fact dim(Ker$\DN$)=$|N|$ 
(we are essentially  in an  instanton background). In terms of an 
orthornormal basis $\Psi^{(N)}_n(x)$ for  Ker $\DN$, an explicit 
expression for such operator is 
given by
\begin{equation}
\PN=\sum_n \Psi^{(N)\dagger}_i (y) \Psi^{(N)}_i (x).
\end{equation}
In the following we shall focus our attention on the  zero modes 
appearing in the fermionic integrations. This issue requires  
some caution in order to get a sensible result \cite{Brazil}.
We begin by  performing  the  change of variables 
\begin{mathletters}
\begin{eqnarray}
&&\psi(x)\to \chi(x)- (\Id-\PN)\circ S^{(N)} \circ (\Id-\PN)\circ\eta_a\\
&&\bar\psi(x)\to \bar \chi(x)-\bar\eta_a\circ (\Id-\PN)\circ S^{(N)}
\circ  (\Id-\PN),
\end{eqnarray}
\end{mathletters}
in the path integral (\ref{Generat1}).
After some cumbersome algebra, our generating functional takes the form
\begin{eqnarray}
\label{Lagrangian1}
&&\!\!\!\!\!\!\!\!\!\!\!\!
Z[\eta,\bar\eta]=\int^1_0 d \alpha_\mu 
\int {\cal D}\bar\chi{\cal D}\chi {\cal D} H_\mu
\exp \Biggl[\int d^2x \det(e) 
\Biggl ( \bar \chi\DN \chi+g^{\mu\nu} H_\mu H_{\nu}\\
&&\!\!\!\!\!\!\!\!\!\!\!\!
-\bar\eta_a\circ (\Id-\PN)\circ S^{(N)}\circ(\Id-\PN)\circ\eta_a
+\bar\eta_a \circ\PN\circ\chi+\bar\chi\circ \PN\circ 
\eta_a\Biggr )\Biggr ].\nonumber
\end{eqnarray}
(Note that the current are now decoupled from the fermion fields 
except the zero modes). The Berezin  integration over the fermionic 
degree of freedom produces
\begin{eqnarray}
\label{Generatdue}
&&\!\!\!\!\!\!\!\!\!\!\!\!\!\!\!\!\!\!\!\!\!\!\!\!\!\!
Z[\eta,\bar\eta]=\int^1_0 d \alpha_\mu 
\int  {\cal D} H_\mu \det ~\! ^\prime (\DN)
\exp \Biggl[\int d^2x \det(e) 
\Biggl (g^{\mu\nu} H_\mu  H_{\nu}\nonumber\\	
&&-\bar\eta_a\circ (\Id-\PN)\circ S^{(N)}\circ(\Id -\PN)\circ\eta_a
\Biggr )\Biggr ]\times \nonumber\\
&&\prod_{n=1}^N\left [
\int dx^2\det(e)\Psi^{(N)\dagger}_n(x)\eta_a(x)\right ]\left [
\int dx^2 \det(e)\bar \eta_a(x)\Psi^{(N)}_n(x)\right];
\end{eqnarray}
$\det ~\! ^\prime (\DN)$ is the (regularized) product of the 
non-vanishing eigenvalues. Our next step will be accomplished with the 
help of the Hodge decomposition  for $H_\mu$, namely
\begin{equation}
\label{Hodge}
H_{\mu}(x)=\partial_\mu \phi_{1}(x)-\eta_{\mu}^{~\nu}
\partial_{\nu}\phi_{2}(x)+{2\pi} h_{\mu}.
\end{equation}
In fact we can use eq.(\ref{Hodge}) and the explicit expression for 
$A^{(1)}_{\mu}$  to obtain
\eq
\DN\!\!=\!\!\exp\!\left [i g \phi_1-\!{3\sigma\over 2}\!-\gamma_5 
\left(g \phi_2+A\sigma-\pi N \lambda \right)\right]\!\!
\hDN\!\! \exp\!\left [-i g \phi_1+\frac{\sigma}{2}\!-\!\gamma_5 \left(g
\phi_2+A\sigma-\pi N\lambda\right)\right ],
\en
with
\eq
\hDN= \hat \gamma^\mu\left (
i\partial_\mu+ N \hat A^{(1)}_\mu+ (N\alpha_{\mu}-g h_{\mu})
\right).
\label{globo}
\en

\noindent
An orthonormal basis  of the zero modes of the 
operator $\hDN$ is related  to the analogous one for $\DN$ by
\begin{equation}
\Psi^{(N)}_i=\exp\left[ig\phi_{1}-\frac{\sigma}{2}+\gamma_{5}
(g\phi_2+A\sigma-\pi N\lambda)\right ] \sum_{j=1}^N C_{ij} \hat\Psi^{(N)}_j
\end{equation}
where we have used a $N\times N$ matrix $C$ to have an orthonormal
basis also for the  null space of $\DN$. We can now compute 
$\det ~\! ^\prime (\DN)$: $\DN$ is a differential elliptic operator acting on 
the sections of a vector-bundle over a compact manifold, and we can 
therefore use the $\zeta$-function definition \cite{Hokky} of the 
determinant to 
get, with a straigthforward application of the technique introduced in 
\cite{Brazil}, the result
\begin{equation}
\det ~\! ^\prime (\DN)=\det ~\! ^\prime {(\hDN)}\exp[-S^{(N)}_{Loc.}]
\mid\det (C)\mid^{-2}.
\label{detto}
\end{equation}

\noindent
We have three different contributions: a ``global'' one, coming from 
the instanton and the flat connections, contained in 
$\det ~\! ^\prime {(\hDk)}$, 
that will be computed by brute force in the following; an highly non-linear 
and non-local part, $\mid\det (C)\mid^{-2}$, linked to the zero-modes 
subtraction; and finally a local bosonic action 
$S^{(N)}_{Loc.}(\phi_1,\phi_2,\lambda,\sigma)$ 
\begin{equation}
S^{(N)}_{Loc.}=S_{Liou.}(\sigma)-
\int d^2x\det(e)\Biggl[{g^2\over 2\pi}\phi_2\triangle\phi_2+
{\pi\over 2}(N\lambda-{A\over \pi}\sigma)\triangle(N\lambda-{A\over \pi}\sigma)
-g\phi_2\triangle(N\lambda-{A\over \pi}\sigma)\Biggr].
\label{actio}
\end{equation}
In order to manage the source-dependent part we introduce the 
operator  $\hat S^{(N)}$ defined  by the equation 
\begin{equation}
\hDN\circ \hat S^{(N)}=\hat S^{(N)}\circ\hDN=\Id-\hPN,
\label{fgrf}
\end{equation}
where $\hPN$ is the self-adjoint projector on the Kernel of 
$\hDN$, one  can prove that \cite{Brazil}
\begin{eqnarray}
\label{Lagrangiancinque}
&&\exp \Biggl[\int d^2x \det(e) \Biggl (-\bar\eta_a\circ
(\Id-\PN)\circ S^{(N)}\circ(\Id -\PN)\circ\eta_a
\Biggr )\Biggr ]\times \nonumber\\
&&\Pi_{i}\left [\int dx^2\det(e)\Psi^{(N)\dagger}_i(x)\eta_a(x) \right ]
\left [\int dx^2\det(e)\bar \eta_a(x)\Psi^{(N)}_i(x)\right]=\nonumber\\
&&\Pi_{i}\left [ 
\int dx^2\det(\hat e)\hat\Psi^{(N)\dagger}_i(x)\eta^\prime_a(x) \right ]\left [
\int dx^2\det(\hat e)\bar \eta^\prime_a(x)\hat\Psi^{(N)}_i(x)\right]\times 
\nonumber\\ 
&&\exp \Biggl[-\int d^2x \det(\hat e) \bar\eta^\prime_a\circ
(\Id-\hPN)\circ \hat S^{(N)}\circ(\Id -\hPN)\circ\eta^\prime_a
\Biggr ]\mid \det (C)\mid^2.
\end{eqnarray}
Here 
\begin{eqnarray}
&&\eta^\prime_a=\exp\left[-ig\phi_{1}-\frac{\sigma}{2}+\gamma_{5}
(g\phi_2+A\sigma-\pi N\lambda)\right ]\eta_a\\
&&\bar \eta^\prime_a=\bar \eta_a
\exp\left[ig\phi_{1}-\frac{\sigma}{2}+\gamma_{5}
(g\phi_2+A\sigma-\pi N\lambda)\right ].
\end{eqnarray}

\noindent
Thus the  generating functional turns out to be:
\begin{eqnarray}
\label{Lagrangiansei}
&&\!\!\!\!Z[\eta,\bar\eta]=2\pi\frac{\exp[-S_{Liou.}(\sigma)]}{\det ~\!^\prime
(\triangle)^{2}}
\int D\phi_1^{\prime}D\phi_2^{\prime} 
\exp \Biggl[\int d^2x\det(e)\Biggl((1+{g^2\over 2\pi})\phi_2\triangle\phi_2+
\phi_1\triangle\phi_1\nonumber\\
&&+{\pi\over 2}(N\lambda-{A\over \pi}\sigma)\triangle(N\lambda-{A\over \pi}
\sigma)-g\phi_2\triangle(N\lambda-{A\over \pi}\sigma)\Biggr)\Biggr]
\int^1_0 d \alpha_\mu 
\int^{+\infty}_{-\infty}\!\!\!\! dh_\mu  
\exp[-\sqrt{\hat g}\hat g^{\mu\nu} h_{\mu} h_{\nu}]\times\nonumber\\
&&\det ~\! ^\prime {(\hDN})\Pi_{i}\left [ 
\int dx^2\det(\hat e)\hat\Psi^{(N)\dagger}_i(x)\eta^\prime_a(x) \right ]
\left [\int dx^2\det(\hat e)\bar \eta^\prime_a(x)\hat\Psi^{(N)}_i(x)\right] 
\\
&&\exp \Biggl[-\int d^2x \det(\hat e) \bar\eta^\prime_a\circ
(\Id-\hPN)\circ \hat S^{(N)}\circ(\Id -\hPN)\circ\eta^\prime_a\Biggr]
\nonumber
\end{eqnarray}

\noindent
The prime on the measure means that the functional integration must be carried 
out only over the non costant modes of $\phi_i$ and 
$2\pi\det^{\prime}(\triangle)^{-1}$ \footnote{As usual the 
symbol $\det^\prime$ means that the zero eigenvalue is excluded.}
is the Jacobian of the change of 
variables from $H_{\mu}$ to $(\phi_1,\phi_2, h_{\mu})$.

\noindent 
Some remarks about  the expression (\ref{Lagrangiansei}) for the generating
functional are now in order. Firstly the highly non local contribution,
due to $\mid \det (C)\mid^2$, disappears from the final expression for the 
generating functional preserving the solvability of the model. Even though 
this result might be  expected  from \cite{Brazil}, it represents a non trivial 
check of our computation. Secondly, from eq. (\ref{Lagrangiansei}), it is 
transparent that the only possibility for a non vanishing partition 
function (${\cal Z}\equiv [0,0]$) is $N=0$. Finally,
from eq. (\ref{Lagrangiansei}) it is straightforward to realize that  
only correlation functions of the form  (\ref{giga}) are different from zero. 
In fact, because of the fixed chirality of the zero modes appering in 
eq. (\ref{Lagrangiansei}), an asymmetry between the number of left and
right current is produced. This, in turn, will entail the asymmetry in 
the Green functions. 

From eq.(\ref{Lagrangiansei}) one can derive all the fermionic (gauge-invariant) 
correlation functions of the theory and by means of appropriate limiting 
procedure the correlators of the relevant composite operators.
\section{The $N=0$ theory}
We start by considering the theory with $N=0$: we find that the central 
charge is modified not only because of the presence of the A-coupling but an 
extra $g^2$-dependence appears through the Thirring interaction. This fact 
is not a priori expected, since , in the minimal-coupled case, 
the central charge is independent of $g^2$. We also derive the explicit 
form of the propagator: this computation could be relevant by itself because, 
at least at our knowledge, does not appear in the literature an explicit 
calculation of the massless Thirring model's two-point function on the torus.
From this computation is not also difficult to obtain the finite-temperature 
propagator, decompactifying the $x^1$-direction.
\subsection{The partition function}
The absence of $B$-coupling simplifies the expression for the generating 
functional (\ref{Lagrangiansei}): no zero-modes are present and the partition 
function does not vanish. Switching off the source terms in 
eq.(\ref{Lagrangiansei}) we have the  following factorization for the partition 
function
\begin{eqnarray} 
\label{path2}
&&{\cal Z}=\exp[-S_{Liou.}(\sigma)]2\pi\det ~\!^\prime(\triangle)^{-1}
\int{\cal D}(\phi_1^{\prime}\phi_2^{\prime}) 
\exp \Biggl[\int d^2x\det(e)\Biggl((1+{g^2\over 2\pi})\phi_2\triangle\phi_2+
\phi_1\triangle\phi_1\nonumber\\
&&+
{A^2\over 2\pi}\sigma\triangle\sigma
+g{A\over \pi}\phi_2\triangle\sigma\Biggr)\Biggr] 
\Biggl[\int^{+\infty}_{-\infty} d h_\mu  
\exp[-\sqrt{\hat g}\hat g^{\mu\nu} h_{\mu} h_{\nu}]
\det ~\! (\hD^{(0)})\Biggr]\nonumber\\
&&\equiv {\cal Z}_{local}
\times {\cal Z}_{global}.
\end{eqnarray} 

\noindent 
From eq. (\ref{path2}) it is also clear that the partition functions splits 
into the product of two factors, which can be computed separately. The former,
${\cal Z}_{local}$,  depends only on the fields $\phi_i(x)$, 
while the latter, ${\cal Z}_{global}$,  depends only on the harmonic form 
$ h_\mu$.

\noindent{\it-- Computation of ${\cal Z}_{local}$}\\
If we introduce the vector notation $\vec{\Phi}(x)\equiv(\phi_1(x), \phi_2(x))$,
${\cal Z}_{local}$  takes the form
\begin{eqnarray}
{\cal Z}_{local}\!\!& =&\!\!
{\rm exp}\left(\!\!-S_{Liou.}(\sigma)+{A^{2}\over 2\pi}\int d^2x\det(e)
\sigma\triangle\sigma\right)\!\!\!\\
&&\int {\cal D} \vec{\Phi}^\prime  
{\rm exp}\left (\int d^2x\det(e)
[-\frac{1}{2}\vec{\Phi}^t {\cal O}
\vec{\Phi}+ \vec{\Phi}^t\vec{J}(x)]
\!\!\right )
2\pi {\det}^\prime (\triangle)\nonumber
\end{eqnarray}
where ${\cal O}$ is the following operator
\begin{equation}
{\cal O}=\triangle \left (
\begin{array}{cc}
\frac{g}{2\pi} &  0 \\ 0 & -2
\end{array} 
\right ), 
\end{equation}
and the source $\vec{J}(x)$ is
\begin{equation}
\vec{J}(x)=({Ag\over \pi}\triangle\sigma\, , \,0)
\end{equation}
The functional integral is gaussian and is easily performed. We obtain
\begin{equation}
\label{Poiu}
{\cal Z}_{local}=2\sqrt{1+{g^2\over 2\pi}}
\exp\left[-\left(1-{12}{A^2\over 1+{g^2\over 2\pi}}\right)S_{Liou.}
(\sigma)\right],
\end{equation}
where the factor $2\sqrt{1+{g^2\over 2\pi}}$ derive from the scaling formula 
for the primed determinant of $\triangle$ \cite{Dette}.

\noindent 
The local part of the partition function confirmes and extends the result of 
\cite{Cangi}: the $A$-coupling changes the value of the central charge of the 
fermionic theory, the coefficient of the Liouville action. A new feature arises 
due to the presence of the current-current interaction, namely the  central 
charge now depends non-trivially on the coupling costant ${g^2\over 2\pi}$.
However  it displays a  pole  for $g^2/2\pi=-1$.  The origin of such a
singularity can be traced back to the well-known loss of unitarity of 
the Thirring model when the coupling constant goes through this value.

\noindent
Finally let us speculate on the possibility of giving a fermionic 
representation of the minimal series by exploiting eq. (\ref{Poiu}).
It is clear that it is possible to get the correct central charge by 
tuning $g$ and $A$. However the question of which screening operators
must be introduced to obtain the correct Green functions is not evident.

\noindent{\it-- Computation of ${\cal Z}_{global}$}

\noindent
We come now to the evaluation of the ${\cal Z}_{global}$, which is  given 
by 
\begin{equation}
{\cal Z}_{global}=\int^{\infty}_{-\infty} d h_\mu 
\exp\left (-4\pi^2\sqrt{\hat g}\hat g^{\mu\nu} h_\mu h_\nu\right)
\det ~\! (\hD^{(0)}).
\end{equation}
The determinant corresponding to the operator $\det ~\! (\hD^{(0)})[ h])$ 
can be computed by a explicit construction  of the associated  $\zeta-$function 
\begin{eqnarray}
{\rm det}[ D ]=\exp\left[-\frac{d}{d s}\zeta_D (s)\right]_{s=0},\\
\zeta_D (s)=\sum_n \lambda_n^{-s}~{\rm deg}(\lambda_n),
\end{eqnarray}
Actually, we shall evaluate the determinant of  $\det ~\! (\hD^{0} )^2$ 
and then we shall extract the square root of such expression. The  operator  
$\hD^{(0)} $ is given by
\begin{equation}
\label{piopo}
-(\hat \triangle- {4\pi i}g \hat g^{\mu\nu} h_{\mu}\hat \nabla_{\nu}-
 {4\pi^2}g^2 \hat g^{\mu\nu} h_\mu  h_\nu ), 
\end{equation}
where $\hat \triangle$ and $\hat \nabla_\mu$ are  the  laplacian and the 
covariant 
derivative in the background metric $\hat g_{\mu\nu}$. Taking into account that
$\hat g_{\mu\nu}$  is flat and independent of $x$,   it is very easy to check that 
the eigenfunctions of this operator are given by
\begin{equation}
\psi_p (x)=e^{-i p_\mu x^\mu} \rho
\end{equation}
where $\rho$  is a constant spinor. The ensuing eigenvalues are 
then obtained by substituting such expression in eq. (\ref{piopo})
\begin{equation}
\lambda_p= \left (p_\mu- {2\pi} g h_\mu\right )
\hat g^{\mu\nu} \left (p_\nu- {2\pi} g h_\nu\right ).
\end{equation}
The degeneracy of each eigenvalue is two. In fact only two  
independent constant spinors $\rho$ exist.

\noindent
The imposition of $(A,A)$ boundary conditions requires that  $p_\mu$
 belongs to the lattice $Z^2+1/2$, namely
\begin{equation}
p_\mu=n_\mu+\frac{1}{2}\ \ \ \ {\rm with}\ n_\mu\in Z^2.
\end{equation}
The associated $\zeta-$function, which defines the determinant, can be now 
easily written down
\begin{equation}
\zeta(s)=2\left(\frac{\tau_2}{2\pi}\right )^{2s}\sum_{ Z^2}
\left [\left (n_1-g h_1+\frac{1}{2}-\tau_1\left ( n_2-g h_2+\frac{1}{2}
\right )\right )^2+\tau_2^2  \left (n_2-g h_2+\frac{1}{2}\right )^2
\right  ]^{-s},
\end{equation}
where we have used the explicit form of the metric to reduce the
expression to this form. The factor $2$ is due to the degeneracy
of the eigenvalue.

\noindent
The explicit computation of $\zeta^\prime (0)$ is not difficult and 
it is given for more general non-hermitian operators in \cite{GS}. Here we shall
only quote the final result, which can be expressed in terms of theta functions
\begin{equation} 
\zeta^\prime(0)\!=\!
-2\log\!\!\left( \frac{1}{|\eta(\tau)|^2}\Theta \left[\!\!
\begin{array}{c}
\displaystyle{g  h_2}\\
\\
\displaystyle{-g  h_1}\\
\end{array}
\!\!\right]\!\!(0,\tau)\
\Theta^* \left[\!\!
\begin{array}{c}
\displaystyle{g  h_2}\\
\\
\displaystyle{-g h_1}\\
\end{array}
\!\!\right]\!\!(0,\tau)\right ) \nonumber
\end{equation}
Thus the determinant for the global  part  is 
$\det ~\! (\hD^{(0)})[ h]=\exp(-\zeta^\prime(0))$.  

\noindent
To determine ${\cal Z}_{local}$ we have now to compute the following 
integral
\begin{equation} 
{\cal Z}_{global}=\int^{\infty}_{-\infty} d h_\mu 
\exp\left (-4\pi^2\sqrt{\hat g}\hat g^{\mu\nu} h_\mu  h_\nu\right)
\frac{1}{|\eta(\tau)|^2}\Theta \left[\!\!
\begin{array}{c}
\displaystyle{g  h_2}\\
\\
\displaystyle{-g  h_1}\\
\end{array}
\!\!\right]\!\!(0,\tau)\
\Theta^* \left[\!\!
\begin{array}{c}
\displaystyle{g  h_2}\\
\\
\displaystyle{-g h_1}\\
\end{array}
\!\!\right]\!\!(0,\tau).\nonumber
\end{equation}

\noindent 
The integral over $h_\mu$ can be performed expanding the theta functions 
as series of exponentials: we end up with a series of gaussian integrations 
whose result can be recast as theta function with characteristic \cite{Mummo}
\begin{equation} 
{\cal Z}_{global}={1\over 2\sqrt{1+{g^2\over 2\pi}}}\Theta(0,\Lambda),
\end{equation}
where the covariance matrix $\Lambda$ is
\begin{equation}
\label{covariance}
\Lambda=\left (\begin{array}{cc}
\tau &  0\\
\\0 & -\bar{\tau}\end{array} \right )+
i {\tau_2 {g^2\over 2\pi}\over 2(1+{g^2\over 2\pi})}\left (\begin{array}{cc}
{g^2\over 2\pi} &  -2-{g^2\over 2\pi}\\
\\-2-{g^2\over 2\pi} & {g^2\over 2\pi}\end{array} \right ),
\end{equation}
and 
\begin{equation}
\Theta(0,\Lambda)=\sum_{\vec{n}\in Z^2}\exp[i\pi \vec{n}\,\Lambda\, \vec{n}].
\end{equation}
Actually to obtain this result we have to impose that ${g^2\over 2\pi}>-1$,
which  
is the usual bound for the unitarity of the Thirring model \cite{Colli}.

\noindent
The final expression for the partition function turns out to be
\begin{equation}
{\cal Z}=\exp\left[-\left(1-{12}{A^2\over 1+{g^2\over 2\pi}}\right)
S_{Liou.}(\sigma)\right]
\Theta(0,\Lambda),
\end{equation}
the only effect of the non-minimal coupling being contained in the Liouville 
sector.

\subsection{The fermionic propagator} 
All the correlation functions of the Thirring model can be constructed in term 
of the building-block propagator, due to the quasi-free character of the 
theory, therefore we limit ourselves to the computation of the 
two-point function.
 
We use the notation:
\eq
\bar{\psi}=(\bar{\psi}_{R},\bar{\psi}_{L}),\,\,\,\,
\psi=\left( \begin{array}{c}\psi_{R}\\ \psi_{L} \end{array}\right).
\label{s81}
\en

\noindent
We can easily derive from the generating functional (\ref{Lagrangiansei}) 
the fermionic propagator:
\eq
\label{pippo31}
<\psi(x)\bar{\psi}(y)>=
\left( \begin{array}{cc}  S_{RR}(x,y)  &  S_{RL}(x,y) \\     
                                      S_{LR}(x,y) &  S_{LL}(x,y)  \end{array}
                                                              \right);
\en
Let us notice that for $N=0$ the additional phase in the propagator disappear.
However in the next section, where we shall address the same question in the 
sector $N=\pm 1$, the presence of the phase depending on $a_\mu$ will play a 
fundamental role. 

\noindent
In absence of zero-modes of the Dirac operator, only $S_{RL}(x,y)$ and
$S_{LR}(x,y)$ are different from zero and we get
\bea
\label{pippo32}
S_{LR,RL}(x,y)&=&{1\over {\cal Z}}\exp\left[
-S_{Liou.}(\sigma)-{1\pm 2A\over 2}\sigma(x)-{1\mp 2A\over 2}\sigma(y)\right]
2\pi \det^{'}(\triangle)
\nonumber\\
&&\int {\cal D}(\phi_i)'\exp\left[-S_{Loc.}[\phi_i]
+ig\left(\phi_1(x)-\phi_1(y)\right)\pm g\left(\phi_2(x)-\phi_2(y)\right)
\right]\nonumber\\
&&\int_{-\infty}^{+\infty} d h_{\mu} S^{G}_{LR,RL}(x,y) \exp\left[
-S_{Glob.}[h]\right].
\label{s82}
\eea

\noindent
We have normalized the propagator to the partition function (as usual) and we 
have introduced $S_{Loc.}[\phi_i]$ and $S_{Glob.}[h]$ that are respectively 
the local part and the global part of the Dirac determinant previously 
computed (we have also included the $H_{\mu}H^{\mu}$ term for convenience)
\bea
&&S_{Loc.}[\phi_i]=-\int d^2x\sqrt{g}\left[(1+{g^2\over 2\pi})
\phi_2\triangle\phi_2
+\phi_1\triangle\phi_1+{A^2\over 2\pi}\sigma\triangle\sigma+
{g A\over \pi}\phi_2\triangle\sigma\right],\nonumber\\
&&S_{Glob.}[h]=-\log|\Theta \left [\begin{array}{c}
g h_2\\-g h_1
\end{array}
\right ](0,\,\tau)|^2
+{4\pi^2\over\tau_2}\left (( h_1-\tau_1 h_2)^2+\tau_2^2 h_{2}^2\right).
\end{eqnarray}
\noindent
$S^{G}_{LR}(x,y)$ is a generalization, to the present 
case, of the well-known Szego kernel ($S^{G}_{RL}(x,y)=S^{G}_{LR}(x,y)^{*}$) 
\begin{equation}
S^{G}_{LR}(x,y)=-{1\over 2\pi }{\Theta\left [\begin{array}{c}
\frac{1}{2}\\\frac{1}{2}
\end{array}
\right ](0,\tau)
\Theta \left [\begin{array}{c}
g h_2\\-g h_1
\end{array}
\right ](z-w,\tau)
\over\Theta\left [\begin{array}{c}
\frac{1}{2}\\-\frac{1}{2}
\end{array}
\right ](z-w,\tau)
\Theta \left [\begin{array}{c}
g h_2\\-g h_1
\end{array}
\right ](0,\tau)}\exp[-2\pi i g( h_1 (x^1-y^1)+
h_2 (x^2-y^2))].
\end{equation}
and we have found it useful to introduce the complex coordinates 
\bea
&&z=(x^2+\tau x^1),\ \ \  w=(y^2+\tau y^1),\nonumber\\
&&\bar{z}=(x^2+\bar\tau x^1),\ \ \ \bar{w}=(y^2+\bar\tau y^1). 
\label{ollo}
\eea

\noindent
The Szego kernels are linked to the Green function of $\hD^{(0)}$: 
\eq
\hD^{(0)}_{x} S^G(x,y)=\frac{I}{\sqrt{\hat g}}\delta^2(x-y) 
\en
where
\eq
S^G(x,y)=
\left( \begin{array}{cc}  0  &  S^G_{RL}(x,y) \\     
                                      S^G_{LR}(x,y) &  0  \end{array}
                                                              \right).
\en 
\noindent
The quadratic integration over $\phi_1,\phi_2$ is performed to obtain
\begin{equation}
\exp\Biggl[\pi{({g^2\over 2\pi})^2\over 1+{g^2\over 2\pi}}
\Biggl( \bar{G}_{\triangle}(x,y)-\bar{G}_{\triangle}(0,0) \Biggr )
\mp A{{g^2\over 2\pi}\over 1+{g^2\over 2\pi}}(\sigma(x)-\sigma(y))
\Biggr],
\end{equation}
\noindent
$\bar{G}_{\triangle}$ being the Green function of the laplacian $\triangle$ 
on the torus of metric $g_{\mu\nu}$, with the zero-modes projected out. 
The integration over the harmonic field $h_{\mu}$ is long but 
straightforward 
leading to final expression for the whole propagator
\bea
\label{proppo1}
&&\!\!\!\!\!\!\!\!\!\!
<\psi(x)\bar{\psi}(y)>=\exp\Biggl[\pi{
({g^2\over 2\pi})^2\over 1+{g^2\over 2\pi}}
\Biggl( \bar{G}_{\triangle}(x,y)-\bar{G}_{\triangle}(0,0) \Biggr )\Biggr]
\times\!\!\!\!\\
&&\!\!\!\!\!\!\!\!\!\!
\left( \begin{array}{cc}
\!\!\!\!\!\!\!\!\!\!\!\!\!\!\!\!\!\!\!\!\!\!\!\!\!\!\!\!\!\!\!
\!\!\!\!\!\!\!\!\!\!\!\!\!\!\!\!\!\!\!\!\!\!\!\!\!\!\!\!\!\!\!
0  & 
\!\!\!\!\!\!\!\!\!\!\!\!\!\!\!\!\!\!\!\!\!\!\!\!\!\!\!\!
\!\!\!\!\!\!\!\!\!\!\!\!\!\!\!\!\!\!\!\!\!\!\!\!\!\!\!\!
\!\!\!\!\!\!\!\!\!\!\!\!\!\!\!\!\!\!\!\!\!\!\!\!\!\!\!
\!\!\!
\hat{S}_{RL}(x,y)
\exp\left (-\frac{1}{2}
\left [\left ({1- 
{2 A\over 1+{g^2\over 2\pi}}}\right)\sigma(x)-
\left ({1+ {2 A\over 1+{g^2\over 2\pi}}
}\right)\sigma(y)\right ]\right )  \\     
 \hat{S}_{RL}^{*}(x,y)\exp\left (-\frac{1}{2}
\left [\left ({1+ 
{2 A\over 1+{g^2\over 2\pi}}}\right)\sigma(x)-
\left ({1- {2 A\over 1+{g^2\over 2\pi}}
}\right)\sigma(y)\right ]\right ) &  0 \end{array}\!\!\!\!\!\right),
\nonumber
\eea
\begin{equation}
\hat{S}_{RL}(x,y)=-{1\over 2\pi }
{\Theta\left [ \begin{array}{c} 
{1\over 2}\\ {1\over 2}
\end{array}
\right ](0,\tau)
\over\Theta \left [\begin{array}{c} 
{1\over 2}\\-{1\over 2}
\end{array}
\right ] (z,\tau)}
\Theta\left [\vec{V}(z)-\vec{V}(w)
\right ](\Lambda,0)\exp\left[{\pi\over 8}{({g^2\over 2\pi})^2
\over 1+{g^2\over 2\pi}}(z-w-\bar{z}+\bar{w})^2\right],
\end{equation}
$\Lambda$ is the covariance matrix eq. (\ref{covariance}) and the vector 
$\vec{V}(z)$ is
\begin{equation}
\vec{V}(z)=\Biggl(z+{1\over 4}{({g^2\over 2\pi})^2\over 1+{g^2\over 2\pi}}
(z-\bar{z}),-{{g^2\over 2\pi}(2+{g^2\over 2\pi})\over 1+{g^2\over 2\pi}}
(z-\bar{z})\Biggr).
\end{equation}

\noindent
To extract the singular behavior in eq. (\ref{proppo1}) we use the expansion 
for $\bar{G}_{\triangle}(x,y)$ as $x\to y$
\begin{equation}
\bar{G}_{\triangle}(x,y)=\bar{G}_{\hat\triangle}(x,y)+
(\lambda(x)+\lambda(y))+ \int\sqrt{g}\lambda {\triangle} \lambda
+O(\delta(x,y)),
\end{equation}
where $\lambda$ was defined in eq. (\ref{nonloc}), 
$\delta(x,y)$ is the geodesic 
distance  and $\bar{G}_{\hat\triangle}(x,y)$ the Green function for the 
flat torus, whose expansion at short distance is 
\begin{equation}
\bar{G}_{\hat\triangle}(x,y)\simeq {1\over 4}\log\left[\mu^2 \hat g_{\mu\nu}
(x^\mu-y^\mu)(x^\nu-y^\nu)\right ]-\frac{1}{4\pi}
\log\left [\frac{4\pi^2\tau_2 |\eta(\tau)|^4}{\mu^2 \hat{V}}\right ]+
O((x-y)^2).
\label{piatta}
\end{equation}
Thus we define the renormalization wave functions
\begin{equation}
\sqrt{Z_{L,R}}=\lim_{x\to y}
\exp\left (\frac{1}{4\pi} 
{({g^2\over 2\pi})^2\over 1+{g^2\over 2\pi}}
 \log\left[\mu^2 \hat g_{\mu\nu}
(x^\mu-y^\mu)(x^\nu-y^\nu)\right ]\right ). 
\end{equation}
In the previous formulae $\mu$ represent the subraction point of our 
renormalization scheme and it corresponds to the usual mass-scale appearing 
in the operator solution of the Thirring model \cite{geppe}.

\noindent
This wave function renormalization is equivalent to the renormalization of the 
fermion field in the flat-space Thirring model \cite{Colli} and it is very much 
expected, being the ultraviolet behaviour independent of  the topology of the 
space-time. 

\noindent
One can check that the propagator is single valued ($z\rightarrow z+1$ and 
$z\rightarrow z+\tau$ are the periodicity in the complex variables) and at 
short distance ($z\rightarrow 0$) the behaviour on the plane is recovered. 
In fact one finds 
\begin{equation}
< \bar \psi(x)\psi(y)>\simeq|z-w|^{-\gamma}
\left( \begin{array}{cc}  0  &  z-w \\     
                                      \bar z-\bar w &  0  \end{array}
                                                              \right).
\end{equation}
where $\gamma=1+1/2(1+g^2/2\pi+1/(1+g^2/2\pi))$. 

\noindent
A further interesting development could be to investigate the 
gravitational dressing of the above propagator (and eventually of 
the four-points function): an explicit A-dependence appears in the 
coefficients of the conformal factor in eq.(\ref{proppo1}) and would be 
interesting to understand its influence on the conformal dimensions of the 
spinor field.

\section{The $N\neq 0$: the chirality violating theory}
Let us now explore how the model changes when we turn on the $B-$coupling.
To have explicit results, we shall only consider in detail the case 
$N=\pm 1$,  the other possible values can be handled in an analogous 
manner. 

\noindent
First of all let us recall that the partition function is now zero. However,
using the generating functional eq.(\ref{Lagrangiansei}) and the fact that for 
$N=\pm 1$  there is only one zero-mode (with respectively different chirality), 
we derive  that  the simplest non vanishing Green function is  the two-point 
one: 
\eq
\label{cippa10}
<\psi(x)\exp\left (i \int_{\gamma_{x0y} } A^{(1)}_\mu dx^\mu\right)
\bar{\psi}(y)>_{N}=
\left( \begin{array}{cc}  S_{RR}(x,y)\delta _{N,-1}&  0 \\     
    0 &  S_{LL}(x,y)\delta _{N,1}  \end{array}\right),
\en
where $S_{LL}(x,y)$ and $S_{RR}(x,y)$ are
\begin{eqnarray} 
\label{kone}
&&S_{LL}(x,y)=2\pi\det ~\!^\prime(\triangle)^{-1}
\Biggl[\int{\cal D}(\phi_1^{\prime}\phi_2^{\prime}) 
\exp[-S^{(1)}_{Loc}+
\int d^2x\det(e)(\phi_1\triangle\phi_1+\phi_2\triangle\phi_2)]
\nonumber\\
&&\exp\left(-ig(\phi_{1}(x)-\phi_{1}(y))-g(\phi_{2}(x)+\phi_{2}(y))
+\pi(\lambda(x)+\lambda(y))-{1+2A\over 2}(\sigma(x)+\sigma(y)\right)\Biggr]
\nonumber\\
&&\Biggl[\int^{1}_{0} d\alpha_\mu
\int^{+\infty}_{-\infty} dh_\mu  
\exp[-g^{\mu\nu}h_{\mu} h_{\nu}]
\det ~\!^\prime (\hDp )\hat{\Psi}^{(0)\dagger}_{1,0}(x)
\hat{\Psi}^{(0)}_{1,0}(y)\exp[iS_{x,y}(x,y;\alpha)]\Biggr]
\end{eqnarray}
\begin{eqnarray} 
\label{ktwo}
&&S_{RR}(x,y)=2\pi\det ~\!^\prime(\triangle)^{-1}
\Biggl[\int{\cal D}(\phi_1^{\prime}\phi_2^{\prime}) \exp[-S^{(-1)}_{Loc}+
\int d^2x\det(e)(\phi_1\triangle\phi_1+\phi_2\triangle\phi_2)]
\nonumber\\
&&\exp\left(-ig(\phi_{1}(x)-\phi_{1}(y))+g(\phi_{2}(x)+\phi_{2}(y))
+\pi(\lambda(x)+\lambda(y))-{1-2A\over 2}(\sigma(x)+\sigma(y))\right)\Biggr]
\nonumber\\
&&\Biggl[\int_{0}^{1} d\alpha_\mu
\int^{+\infty}_{-\infty} d h_\mu  
\exp[-g^{\mu\nu} h_{\mu} h_{\nu}]
\det ~\!^\prime (\hDm)
\hat{\Psi}^{(0)\dagger}_{-1,0}(x)\hat{\Psi}^{(0)}_{-1,0}(y)
\exp[-iS_{x,y}(x,y;\alpha)]\Biggr].
\end{eqnarray}
\noindent
Here $S^{(\pm 1)}_{Loc}$ is the same action in eq.(\ref{actio}) for $N=\pm 1$, 
$\exp[iS_{x,y}(x,y;\alpha)]$ is the phase factor introduced in 
eq.(\ref{cippa10}) to preserve the global definition of the Green'functions and 
$\hat{\Psi}^{(0))}_{\pm 1,0}(x)$ are the zero modes of the Dirac operator 
$\hDpm$, explicitly derived in the Appendix. 

\noindent 
By comparing this expression with eqs. (\ref{pippo31}) and (\ref{pippo32}), 
the main difference with the $N=0$ theory is manifest: the tensorial 
structure is completely different, namely we have a diagonal propagator 
instead of an antidiagonal one. Moreover there is an explicit dependence 
from the field $\lambda$, that in the previous theory ($N=0$) was introduced 
only through the renormalization. 

\noindent
First of all we compute the global phase 
\eq
S_{x,y}(x,y;\alpha)=\int_{\gamma_{x0y}} {A}^{(1)}_{\mu} dx^{\mu}.
\en
A direct computation is too lengthy, so we proceed by a means of a trick.
We introduce the straight line $r_{xy}$ connecting  $x$ with  $y$ and the 
closed circuit $C$ formed by $\gamma_{xOy}$ and $r_{xy}$; then, with the help
of these two paths, we rewrite the phase as follow
\eq
S_{x,y}(x,y;\alpha)=\oint_{C} {A}^{(1)}_{\mu} dx^{\mu}
+\int_{r_{xy}} {A}^{(1)}_{\mu} dx^{\mu}.
\en
The first term is evaluated by using the Stokes theorem and it gives  the area
enclosed by $C$. The second one can be almost explictly computed by means of the
explicit expression for $A^{(1)}_\mu$. At the end, combining the two results, 
we have
\begin{eqnarray}
&&S_{x,y}(x,y;\alpha)=Area_C+\int_0^1 dt \eta_{\mu}^{~\nu}(y^\mu-x^\mu)
\partial_\nu
\lambda( t y+(1-t)x )-\\
&&i\left[\bar{\alpha}(\bar{z}-\bar{w})+\alpha(z-w)\right]
+i{\pi\over 2\tau_2}(z\bar{w}-\bar{z}w),
\label{pase}\nonumber
\end{eqnarray}
where we have introduced the notation
\bea
&&\alpha={\pi\over i\tau_2}(\alpha_1-\bar{\tau}\alpha_2)\nonumber\\
&& h={\pi\over i\tau_2}(h_1-\bar{\tau}h_2)\nonumber\\
&&\bar{\alpha}=\alpha^*\,\,\,\, , \,\,\, \bar{h}=h^*
\eea
and the  holomorphic coordinate $z$ and $w$.

\noindent
The computation of (\ref{kone}) and (\ref{ktwo}) is again performed 
in two different steps, one for the quadratic 
integration over the local  degrees of freedoom and the other for the more 
involved finite-dimensional integrals over $\alpha$ and $h$.

\noindent
The quadratic integration over $\phi_1,\phi_2$ is standard: 
it gives a contribution to $S_{LL}(x,y)$ equal to
\bea
&&S^{Loc.}_{LL}(x,y)=2\sqrt{1+{g^2\over 2\pi}}\exp\Biggl[-S_{Liou.}(\sigma)+
{\pi\over 2}{1\over 1+{g^2\over \pi}}\Biggl
(\lambda(x)-{A\over \pi}\sigma(x)+\nonumber\\
&&\lambda(y)-{A\over \pi}\sigma(y)\Biggr)
-{1\over 2}(\sigma(x)+\sigma(y))\Biggr]
\exp\Biggl[\pi\Biggl(1+{g^2\over 2\pi}-
{1\over 1+{g^2\over 2\pi}}\Biggr)\bar{G}_{\triangle}(x,y)-\nonumber\\
&& \pi{({g^2\over 2\pi})^2\over 1+{g^2\over 2\pi}}
\bar{G}_{\triangle}(x,x)+{\pi\over 2}{1\over 1+{g^2\over 2\pi}}\int d^2x
\det(e)(\lambda-{A\sigma\over \pi})
\triangle (\lambda-{A\sigma\over \pi})\Biggr]
\eea
while $S^{Loc.}_{RR}(x,y)$ is obtained by changing $A$ in $-A$. One can check 
that the ultraviolet renormalization costant $Z_{L,R}$ previously introduced 
automatically makes the correlation function finite.

\noindent
The computation of the global part $S^{Glob.}_{LL}(x,y),S^{Glob.}_{RR}(x,y)$ is 
more involved: first of all we need \\$\det^{\prime}(\hDpm [\alpha, h])$. 
The eigenvalues and their degeneration are derived in the Appendix:
\eq
(\lambda_{n})^{2}=2n|\Phi|\,\,\, , \,\,\, |\Phi|={2\pi\over \hat{V}}
\en
with degeneration 2. Using $\zeta-$function definition we get
\eq
\det^{\prime}(\hDpm [\alpha,h])=({\pi\over \Phi})^{{\Phi\over 4\pi}}.
\en
We notice that this determinant does not depend $\alpha_\mu$ and $ h_\mu$: 
this is no longer true for the zero-modes as one may notice looking at their 
explicit functional form presented in the Appendix. 

\noindent
The integral we have to perform is (we restrict ourselves for the moment to 
$N=1$)
\bea
&&\!\!\!\!\!\!\!\!\!\!\!
S^{Glob.}_{LL}(x,y)=\sqrt{2\over \tau_2}\int_{0}^{1}d\alpha_1
d\alpha_2
\exp\Biggl[-{4\pi^2\over \tau_2}(\alpha_1,\alpha_2)
\left( \begin{array}{cc}  1&  -\tau_1 \\     
    -\tau_1 &  |\tau|^2 \end{array}\right)
\left( \begin{array}{c}  \alpha_1\\     
    \alpha_2 \end{array}\right)\Biggr]\\
&&\!\!\!\!\!\!\!\!\!\!\!
\int_{-\infty}^{+\infty}\!\!\!\!\!\!
d h_1 d h_2
\exp\Biggl[-{4\pi^2\over \tau_2}(h_1, h_2)
\left( \begin{array}{cc}  1&  -\tau_1 \\     
    -\tau_1 &  |\tau|^2 \end{array}\right)
\left( \begin{array}{c}  h_1\\     
    h_2 \end{array}\right)-{8\pi^2\over \tau_2}(h_1,h_2)
\left( \begin{array}{cc}  1&  -\tau_1 \\     
    -\tau_1 &  |\tau|^2 \end{array}\right)
\left( \begin{array}{c}  \alpha_1\\     
    \alpha_2 \end{array}\right)
\Biggr]\nonumber\\
&&\!\!\!\!\!\!\!\!\!\!\
\exp\Biggl[-{\tau_2\over 2\pi}(g h+g\bar{ h})^2+
({\tau_2\over 2\pi}z+ig\bar{ h})(z-\bar{z})
-({\tau_2\over 2\pi}\bar{w}-ig h)(w-\bar{w})\Biggr]
\!\!\Theta^{*}\!\!\!\left [\!\!\begin{array}{c}
                   {1\over 2} \\
          {i\over\pi}g h\tau_2+{1\over 2}
\end{array}\!\!\right ]
(-\bar{z},-\bar{\tau})\nonumber\\
&&\!\!\!\!\!\!\!\!\!\!\
\Theta\left [\begin{array}{c}
{1\over 2}\\
{i\over\pi}g h\tau_2+{1\over 2}
\end{array}\right ]
(-\bar{w},-\bar{\tau})
\exp\Biggl[-i\left[\bar{\alpha}(\bar{z}-\bar{w})+\alpha(z-w)\right]
+i{\pi\over 2\tau_2}(z\bar{w}-\bar{z}w)\Biggr]\nonumber
\eea
where we have used the explicit expression for the zero-mode eq.(\ref{Pumpk}), 
obtained in the Appendix. We have obviously included in the previous 
integration the $\alpha_\mu$-dependent piece in the phase factor 
eq. (\ref{pase}). Expanding the $\Theta$-functions as a series of 
exponential the integral over ${ h_1, h_2}$ is of gaussian type, and 
after some boring but straightforward computations we remain with
\bea
&&\!\!\!\!{(1+{g^2\over
2\pi})^{-1}\over 2\pi\tau_2  }
\sum_{m,n=-\infty}^{+\infty}\int_{0}^{1}d\alpha_1 d\alpha_2
\exp\Biggl[{1\over 2}\vec{h}\,\tilde{\Lambda}\,\vec{h}
-{2\pi\tau_2\over 1+{g^2\over
2\pi}}{\alpha_2}^2-2\pi i(z-\bar{w})\alpha_2
-{g^2\tau_2\over 1+{g^2\over2\pi}}(h_1+h_2)\alpha_2\nonumber\\
&&\!\!\!\!-2\pi i(m-n)\alpha_1\Biggr]
\exp\Biggl[i\pi\tau(n+{1\over 2})^2-i\pi\bar{\tau}(m+{1\over 2})^2
-2\pi i\Biggl((n+{1\over 2})\tau-(m+{1\over 2})\bar{\tau}\Biggr)
+\\
&&\!\!\!\!2\pi i(n+{1\over 2})(z+{1\over 2})
-2\pi i(m+{1\over 2})(\bar{w}+{1\over 2})\Biggr]
\exp\Biggl[
{\pi\over 2\tau_2}\Biggl(
z(z-\bar{z})-\bar{w}(w-\bar{w})+\bar{z}w-\bar{w}z\Biggr)\Biggr]
\nonumber
\eea
the matrix $\tilde{\Lambda}$ being
\eq
-{\pi\over 2}{\tau_2 {g^2\over
2\pi}\over 1+{g^2\over
2\pi}}\left( \begin{array}{cc}  {g^2\over
2\pi}&  -2-{g^2\over
2\pi}\\     
    -2-{g^2\over
2\pi}&  {g^2\over
2\pi}\end{array}\right)
\en
and the vector $\vec{h}$ is 
$(n+{1\over 2}-{i\over 2\tau_2}(z-\bar{z}),
m+{1\over 2}-{i\over 2\tau_2}(w-\bar{w})$. The integration over $\alpha_1$ 
can now be performed leading to a $\delta_{m,n}$ in the sum: we remark that 
without the introduction of the phase factor this nice integration cannot 
be done, because the integral would be quadratic in $\alpha_1$. Another magic 
of the phase factor is that now the exponential factor can be recast in a 
way that the sum over $n$ simply changes the integration region from $(0,1)$ 
to $(-\infty,+\infty)$, allowing another easy gaussian integration. The final 
expression is very simple:
\eq
S^{Glob.}_{LL}(x,y)=
{1\over 2\pi\hat{V}}\exp\left[-{\pi\over2\tau_2}(1+{g^2\over 2\pi})
(z-w)(\bar{z}-\bar{w})\right],
\en
and the same result applies for $S^{Glob.}_{RR}(x,y)$.

\noindent
The final expression for the full (gauge-invariant) propagator is:
\eq
S_{LL,RR}(x,y)=S^{Loc.}_{LL,RR}(x,y)S^{Glob.}_{LL,RR}(x,y)\exp\pm i\Biggl[
Area_C + \int_0^1 dt \eta_{\mu}^{~\nu}(y^\mu-x^\mu)
\partial_\nu\lambda( t y+(1-t)x )\Biggr],
\en
where the translational invariance is broken only by the geometrical factors 
linked to the background conformal factor. The short distance behaviour 
of this propagator is completly determined by its local part, no contribution 
coming from the integration over the harmonic field:
\eq
< \bar \psi(x)_{L,R}\psi(y)_{L,R}>\simeq|z-w|^{-\delta}
\end{equation}
where $\delta=1/2(1+g^2/2\pi-1/(1+g^2/2\pi))$. This must be compared with the 
analogous scaling in the $N=0$ theory: it is different showing how the 
volume-interaction has changed the scaling dimension of the Thirring fermion. 
The coefficient $\delta$ is always positive in the unitarity range 
(${g^2\over 2\pi}>-1$), except for the free-fermion case where it vanishes. 
We do not have an explanation for such a behaviour and could be interesting 
to compare our results with the scaling laws obtained in \cite{Cabra}, where 
the Thirring model in topological non-trivial background was studied. 

\noindent
As we have promised in the introduction, we present a nice application of 
finite-temperature quantum field theory to our model: for $N=\pm 1$ the 
theory develops a fermionic condensate at $T\neq 0$ temperature. 
From the explicit expression of the propagator is, in fact, 
possible to extract, 
by means of a coincidence limit, the vacuum expectation value of the composite 
operator $\langle \bar\psi(x)\psi(x)
\rangle=\langle \bar\psi_{L}(x)\psi_{L}(x)+\bar\psi_{R}(x)\psi_{R}(x)
\rangle$. For the finite-temperature interpretation we take $\sigma=0$ 
(flat-space), we define the temperature $T$ from $\tau=i{1\over TL}$ and then 
we perform the thermodynamical limit $L\rightarrow \infty$ (we have to 
restore in the correlators the factor $L$, measuring the size of the torus). 
In the limiting procedure a new divergence arises, linked to the composite 
operator $\bar\psi(x)\psi(x)$, that must be subtracted at the same scale $\mu$ 
where the wave function renormalization was defined. Using the expansion 
for the scalar Green function on the flat-torus, described in eq (\ref{piatta})
, and 
letting $L\rightarrow \infty$ we get:
\eq
\langle \bar\psi(x)\psi(x)\rangle_{N=\pm 1}=
{1\over 2\pi^2}\sqrt{1+{g^2\over 2\pi}}
T\Biggl({\mu\over T}\Biggr)^{{{g^2\over 2\pi}\over 1+{g^2\over 2\pi}}}.
\end{equation}

We notice that in the region of unitarity the zero-temperature limit is 
regular, leading to a vanishing condensate. The scaling exponent of the 
temperature is ${1 \over 1+{g^2\over 2\pi}}$ that reduces to $1$ in the case 
of non self-interacting fermions. The result, that might be somewhat 
unexpected being a sort of gravitational mechanism for vacuum condensate, is 
easily understood exploiting the analogy of the volume-potential with 
the instantonic configurations of the two dimensional Maxwell theory, that 
generate $<\bar\psi(x)\psi(x)>\neq 0$ in the Schwinger model.

\section{Conclusions}
In conclusion we have thoroughly studied on the two-sphere $S^2$ and the 
torus $T^2$ a fermionic system coupled in a non-minimal way to gravity: the 
strenght of the new interactions, introduced by Cangemi and Jackiw in 
ref.\cite{Cangi}, must be quantized in order to preserve global properties 
of the theory. The non-minimal coupling with the curvature and the unusual 
interaction with the volume form drastically change the dynamics 
of the masseless abelian Thirring model. In particular the effective action 
for the Liouville mode, obtained by integrating out the fermionic degrees of 
freedoom, displays some new features: the central charge is improved, depending 
also on the Thirring coupling costant, and a new non-local functional of the 
conformal factor appears. It could be very interesting to explore the dynamics 
of this effective action, representing a very natural extension of the usual 
Polyakov gravity, even if the task does not seem very simple. 
On the other hands the presence of zero-modes in the Dirac operator, 
that can be generated by the coupling with the volume form, imposes 
chiral selection rules on the fermionic Green functions. On a general 
Riemann surface (and, in particular, on the torus) there is a family of 
gauge-inequivalent volume-potentials: we have explicitly integrated over 
this moduli space in order to preserve the global definition and the 
traslational invariance of the two-point fermionic function, that has been 
carefully computed. In absence of $B-$coupling the propagator 
presents the same singularity 
structure of the theory on the plane, as one could expect. A peculiar 
effect of the volume coupling is the appearence of a diagonal part for 
these propagator and, eventually, of a non-vanishing fermionic condensate: it 
represents a sort of non-pertubative gravitational modification of the vacuum
structure of the original fermionic theory. It is also possible to analyze 
our results in the spirit of finite temperature quantum 
field theory: anti-periodic boundary conditions for fermions on the flat torus 
allow for the usual finite-temperature interpretation of our correlation 
functions, that therefore encode the thermal dependence of the Thirring model 
dynamics. This could be the subject of further investigations.

\section*{Acknowledgement}
We thank Prof. Roman Jackiw for having read the manuscript and useful 
comments on the subject.

\appendix
\section{Appendix}
In this appendix we derive the spectrum and the eigenfunctions of the 
operator $(\hDN)^2 [\alpha, h]$. 

\noindent
Using the explicit expression (\ref{instantont}) for the $\hat{A}^{(N)}_{\mu}$ 
potential 
we write the eigenvalue problem for $(\hDN)^2$ as:
\begin{equation}
[-\hat{g}_{\mu\nu}\hat{D}_{\mu}\hat{D}_{\nu}-\Phi\gamma_5]\hat{\Psi}^{(n)}=
\lambda_{n}\hat{\Psi}^{(n)},
\label{app1}
\end{equation}
where we have introduced the topological flux $\Phi={2\pi N\over \hat{V}}$ 
and the covariant derivative
\bea
&&\hat{D}_{\mu}=\partial_{\mu}+\Omega_{\mu}
\nonumber\\
&& \Omega_{\mu}=-\Phi\eta_{\mu\nu}x^{\nu}-{2\pi}(k\alpha_{\mu}-
g h_{\mu}).
\eea
In order to solve the eigenvalue equation we find useful to work with the 
holomorphic coordinates $z,\bar{z}$ of eq.(\ref{ollo}) and to define
\bea
&&\Omega=-i{\Phi L\over 4}\bar{z}+\gamma\nonumber\\
&&\bar{\Omega}=i{\Phi  \over 4}z+\bar{\gamma}\nonumber\\
&&\gamma={1\over 2i\tau_{2}}\Biggl[k\alpha_1-g h_1-\bar{\tau}(k\alpha_2-
g h_2)\Biggr]\nonumber\\
&&\bar{\gamma}=\gamma^*.
\eea

\noindent
Eq. (\ref{app1}) can be rewritten as
\bea
&&[b^{\dagger}b+{1+\gamma_5\over 2}]\hat{\Psi}^{(n)}=\hat{\lambda}_{n}
\hat{\Psi}^{(n)}
\,\,\,\,\,\,\,\,\,\, k>0 \nonumber\\
&&[b\,b^{\dagger}+{1-\gamma_5\over 2}]\hat{\Psi}^{(n)}=\hat{\lambda}_{n}
\hat{\Psi}^{(n)}
\,\,\,\,\,\,\,\,\,\, k<0,
\eea
where the operators
\bea
&&b=i\,\sqrt{{2\over  |\Phi|}}(\partial_{z}+i\Omega)\nonumber\\
&&b^{\dagger}=i\,\sqrt{{2\over |\Phi|}}(\partial_{\bar{z}}+i\bar{\Omega})
\eea
satisfy the algebra
\bea
&&[b,b^{\dagger}]=1 \,\,\,\,\,\,\,\,\,\, k>0\nonumber\\
&&[b,b^{\dagger}]=-1 \,\,\,\,\,\,\,\,\, k<0,
\label{app2}
\eea
and $\lambda_{n}=2|\Phi|\hat{\lambda}_{n}$.

\noindent
The algebra (\ref{app2}) immediatly leads to the spectrum
\eq
\lambda_{n}=2n|\Phi|.
\en
Moreover if $k$ is the degeneration of the null eigenvalue we recognize 
(due to the presence of the projector ${1\pm\gamma_5\over 2}$) that $2k$ is 
the degeneration of the non-vanishing ones. For $N>0$ zero-modes have negative 
chirality: all the tower of eigenfunctions 
are simply obtained (up normalizations) acting with $b^{\dagger}$:
\bea
&&\hat{\Psi}_{i}^{(n+1,+)}=
(b^{\dagger})^n\left( \begin{array}{c} \hat{\Psi}^{(0)}_{i}\\  
  0\end{array} \right),\nonumber\\
&&\hat{\Psi}_{i}^{(n,-)}=(b^{\dagger})^n\left( \begin{array}{c} 0\\   
 \hat{\Psi}^{(0)}_{i}\end{array} \right),
\eea
$\hat{\Psi}^{(0)}_{l}$ being solutions of the ground state equation
\eq
b\hat{\Psi}^{(0)}_{l}=0.
\label{app3}
\en
For $N<0$ the chirality is reversed 
and the roles of $b$ and $b^{\dagger}$ are interchanged.

\noindent
We are therefore left with solving eq.(\ref{app3}), on the space of functions 
that satisfy the boundary conditions 
\begin{mathletters}
\begin{eqnarray}
&&\psi(z+\tau ,\bar z +\bar \tau)=\exp \left ( i\pi -\frac{\pi N}
{2\tau_2}
(\bar \tau z-\bar z \tau)\right)\psi (z,\bar z), \\
&&\psi(z+1,\bar z +1)=\exp \left (i \pi -\frac{\pi N}{2\tau_2}
( z-\bar z)\right)\psi (z,\bar z),
\label{bondo}
\end{eqnarray}
\end{mathletters}  
where the antiperiodic boundary conditions has been taken into account.
A general solution of eq.(\ref{app3}) is
\eq
\hat{\Psi}^{(0)}(z,\bar{z})=\exp\left[-({N\pi\over 2 \tau_2}\bar{z}+i\gamma)
(z-\bar{z})\right]\hat{\Psi}^{(0)}(\bar{z})
\en
and the boundary conditions on the anti-holomorphic function 
$\hat{\Psi}^{(0)}(\bar{z})$ are
\bea
&&\hat{\Psi}^{(0)}(\bar{z}+1)=-\hat{\Psi}^{(0)}(\bar{z}),\nonumber\\
&&\hat{\Psi}^{(0)}(\bar{z}+\bar{\tau})=-\exp[2\pi i N\bar{z}+i\pi N\bar{\tau}-
2\tau_{2}\gamma]\hat{\Psi}^{(0)}(\bar{z}).
\eea
One can explicitly verify that
\eq
\hat{\Psi}^{(0)}_{N,l}(\bar{z})=
\Theta\left [\begin{array}{c}
{l\over 2N}\\
-{i\pi}\tau_2\gamma_z+{1\over2}
\end{array}
\right ]
(-N\bar{z},-N\bar{\tau})
\en
with $l=0,1,...|N|-1$ have the required properties: they are exactly $|k|$, as 
stated by the index theorem \cite{Atty}. Moreover it is not difficult to show the 
orthogonality of the zero modes $\hat{\Psi}^{(0)}_{k,l}(z,\bar{z})$: 
they can also 
be normalized to unity to give an ortonormal basis for the kernel of $\hDN$ 
($N>0$)
\eq
\hat{\Psi}^{(0)}_{N,l}(z,\bar{z})=({2\over \tau_{2}})^{1\over 4}
\exp[-{\tau_{2}\over 4\pi}(\gamma+\bar{\gamma})^2]\exp[-({N\pi\over 2 \tau_2}
\bar{z}+i\gamma)(z-\bar{z})]
\Theta\left [\begin{array}{c}
{l\over 2N}\\
-{i\pi}\tau_2\gamma_z+{1\over2}
\end{array}\right ](-N\bar{z},-N\bar{\tau}).
\label{Pumpk}
\en
For $N<0$ we get along the same lines
\eq
\hat{\Psi}^{(0)}_{N,l}(z,\bar{z})=({2\over \tau_{2}})^{1\over 4}
\exp[-{\tau_{2}\over 4\pi}(\gamma+\bar\gamma)^2]
\exp[-({N\pi\over 2 \tau_2}z-i\bar\gamma)
(z-\bar{z})]
\Theta\left [ \begin{array}{c}
{l\over 2N}\\
{i\pi}\tau_2\bar\gamma-{1\over2}
\end{array}
\right ]
(-Nz,-N\tau).
\en


\newpage

\begin{thebibliography}{100}
\bibitem{Topi}{N.K.Nielsen and B.Schroer, {\it Nucl. Phys.} {\bf B127}, 493 
(1977);\\ 
K.D.Rothe and J.A.Swieca, {\it Ann. Phys. (N.Y.)} {\bf 117}, 382 (1979);\\
M.Hortacsu, K.D.Rothe and B.Schroer, {\it Phys. Rev. D} {\bf 20}, 3203 (1979);
\\
I.Sachs and A.Wipf, {\it Helv. Phys. Acta} {\bf 65}, 653 (1992);\\
A.Smilga, {\it  Phys. Rev. D} {\bf 49}, 6836 (1994);\\
D.J.Gross and A.Matytsin, {\it Nucl. Phys.} {\bf B429}, 50 (1994).}
\bibitem{Geomi} {D.Z.Freedmann and K.Pilch, {\it  Phys.Lett.} {\bf B213}, 331 
(1988); {\it Ann. Phys. (N.Y.)} {\bf 192}, 331 (1989);
F.Ferrari,  {\it Nucl. Phys.} {\bf B439}, 692 (1995).}
\bibitem{Polli}{A.M.Polyakov, {\it  Mod. Phys. Lett.} {\bf A2}, 893 (1987);\\
V.G.Knizhnik, A.M.Polyakov and A.B.Zamolodchikov, {\it Mod. Phys. Lett.}
{\bf A3}, 819 (1988);\\
F.David, {\it Mod. Phys. Lett.} {\bf A3}, 1615 (1988);\\
J.Distler and H.Kawai, {\it Nucl. Phys.} {\bf B321}, 509 (1989).}
\bibitem{Giacchi}{
C.Teitelboim, {\it Phys. Lett.} {\bf B126}, 41 (1983);\\
R. Jackiw, {\it Nucl. Phys.} {\bf B252}, 343 (1985)\\
C.Callan, S.B.Giddings, J.A.Harvey and A.Strominger, {\it Phys. Rev. D}
{\bf 45}, 1005 (1992);\\
G.Grignani and G.Nardelli, {\it  Nucl. Phys.} {\bf B412}, 320 (1994).}
\bibitem{Cangi}{D.Cangemi and R.Jackiw, {\it  Ann. Phys. (N.Y.)} {\bf 225}, 229 
(1993).} 
\bibitem{Cangi1}{D.Cangemi and R.Jackiw, {\it Phys. Lett.} {\bf B225}, 229 (1993).} 
\bibitem{Globi}{R.Jackiw,{\it Topological investigation of quantized gauge 
theories}, in {\it Current algebra and Anomalies}, Princeton Series in Physics, 
Princeton (New Jersey) (1985).}
\bibitem{Raggio}{R.Rajaraman, {\it Solitons and Instantons}, North-Holland, 
Amsterdam, (1982).}
\bibitem{Frie}{D. Friedmann and K. Pilch, {\it Phys. Lett. } {\bf B213}, 331 
(1988); {\it Ann. of Phys. (N.Y.)} {\bf 192}, 401.}
\bibitem{Gri}{A.Bassetto and L.Griguolo, {\it Nucl. Phys.} {\bf B439}, 327 
(1995).}
\bibitem{Atty}{M.F.Atiyah and I.M.Singer, {\it Ann. Math. {\bf 87}}, 485, 546, 
(1969); {\bf 93}, 1, 119, 139 (1971).}
\bibitem{Doky}{L.Alvarez-Gaum\'e, G.Moore and C.Vafa, {\it  Comm. Math. Phys.}
{\bf 106}, 1 (1986).}
\bibitem{Libro}{S. Kobayashi, K. Nomizu, {\it Foundation of differential 
geometry}, Interscience (1963).} 
\bibitem{Bardi}{K.Bardakci and M.Crescimanno, {\it  Nucl. Phys.} {\bf B313}, 
269 (1989).} 
\bibitem{Thir}{W. Thirring, {\it Ann. of Phys. (N.Y.)} {bf 3}, 91 (1958)}
\bibitem{Baggi}{J. Bagger, D. Nemeschansky, N. Seiberg and S. Yankielowicz,
{\it Nucl. Phys.} {\bf B289}, 53 (1987).} 
\bibitem{Destri}{C. Destri and J. J. de Vega, {\it  Phys. Lett.} {\bf B223}, 
365 (1989).}
\bibitem{Bote}{L. Botelho, {\it  Europhys. Lett. } {\bf 11}, 313 (1990)}.
\bibitem{Wippi}{I. Sachs and A. Wipf, {\it Genralized Thirring Models}, hepth
9508142}
\bibitem{Hokky}{S.W. Hawking, {\it Comm. Math. Phys.} {\bf 55}, 133 (1977).}
\bibitem{Brazil}{S.A.Dias and M.T.Thomas, {\it Phys. Rev. D.} {\bf 44}, 1672  
(1991).}
\bibitem{Dette}{S.Blau, M.Visser and A.Wipf, {\it Int. J. Mod. Phys.}
{\bf A4}, 1467 (1989).}
\bibitem{GS}{L.Griguolo and D.Seminara, {\it The generalized chiral Schwinger model 
on the torus: an investigation on geometrical, confinement and finite 
temperature properties}, in preparation.} 
\bibitem{Mummo}{D.Mumford, {\it  Tata lectures on theta functions}, 
Birkhauser, Basel 1983.} 
\bibitem{Colli}{S. Coleman, {\it Phys. Rev. D} {\bf 11}, 2088 (1975).} 
\bibitem{geppe}{K. Johnson, {\it Nuovo Cim.} {\bf 20}, 773 (1964);\\
B. Klaiber, {\it "Lectures note in Phys. XA"}, Gordon and Breach, New York (1968).}
\bibitem{Cabra}{D. Cabra, E. Moreno and C. Naon, {\it Nucl. Phys} {\bf B424}, 
567 (1995)          }
\end{thebibliography}
\end{document}